\documentclass[twocolumn]{aastex61}
\pdfoutput=1
\usepackage{amsmath,amstext}
\usepackage[T1]{fontenc}
\usepackage[figure,figure*]{hypcap}
\usepackage{tikz}
\usepackage{CJK}
\usetikzlibrary{shapes,arrows}

\let\oldAA\AA
\renewcommand{\AA}{\text{\normalfont\oldAA}}

\newcommand{\logg}{$\log{g}$ }

\newcommand{\Te} {$T_{\rm eff}~$}

\shorttitle{Modeling Circumstellar Features of WD1145+017}
\shortauthors{Fortin-Archambault et al.}

\begin{document}
\begin{CJK}{UTF8}{gbsn}
\title{Modeling of the Variable Circumstellar Absorption Features of WD 1145+017
}

\author{M. Fortin-Archambault}
\affiliation{Institut de Recherche sur les Exoplan\`etes (iREx) and D\'epartement de Physique, Universit\'e de Montr\'eal, Montr\'eal, QC H3C 3J7, Canada, maude@astro.umontreal.ca, dufourpa@astro.umontreal.ca
QC H3C 3J7, Canada; maude@astro.umontreal.ca, dufourpa@astro.umontreal.ca}
\author{P. Dufour}
\affiliation{Institut de Recherche sur les Exoplan\`etes (iREx) and D\'epartement de Physique, Universit\'e de Montr\'eal, Montr\'eal, QC H3C 3J7, Canada, maude@astro.umontreal.ca, dufourpa@astro.umontreal.ca
QC H3C 3J7, Canada; maude@astro.umontreal.ca, dufourpa@astro.umontreal.ca}
\author{S. Xu (许\CJKfamily{bsmi}偲\CJKfamily{gbsn}艺)}
\affiliation{Gemini Observatory, 670 N. A'ohoku Place, Hilo, HI 86720}

\begin{abstract}
We present an eccentric precessing gas disk model designed to study the variable circumstellar absorption features detected for WD 1145+017, a metal polluted white dwarf with an actively disintegrating asteroid around it. This model, inspired by one recently proposed by Cauley et al., calculates explicitly the gas opacity for any predetermined physical conditions in the disk, predicting the strength and shape of all absorption features, from the UV to the optical, at any given phase of the precession cycle. The successes and failures of this simple model provide valuable insight on the physical characteristics of the gas surrounding the star, notably its composition, temperature and density. This eccentric disk model also highlights the need for supplementary components, most likely circular rings, in order to explain the presence of zero velocity absorption as well as highly ionized Si IV lines. We find that a precession period of $4.6\pm0.3$ yrs can successfully reproduce the shape of the velocity profile observed at most epochs from April 2015 to January 2018, although minor discrepancies at certain times indicate that the assumed geometric configuration may not be optimal yet. Finally, we show that our model can quantitatively explain the change in morphology of the circumstellar features during transiting events.

\end{abstract}

\section{Introduction}
 
It has become widely accepted now that the source of the heavy elements observed in the photosphere of some white dwarfs is accretion from tidally disrupted planets or planetesimals \citep[see][and references therein]{JuraYoung2014}. The discovery by \citet{Vanderburg2015} of an ongoing disintegration event around the white dwarf WD 1145+017 has recently solidified the confidence in this scenario, opening at the same time a new window onto our understanding of this process. 

This unique system shows some remarkable characteristics. Six stable periods between 4.5 and 5 hours were detected in the original K2 light curve \citep{Vanderburg2015}. These periods have been interpreted as being the signature of several smaller objects that have broken off from one main body orbiting the star. These fragments are thought to drift away from the main orbit and slowly disintegrate into dust and gas before being accreted onto the surface of the star \citep{Rappaport2016, Veras2017}. The system is evolving constantly, with changes in the light curve on timescales ranging from minutes to months \citep{Gaensicke2016, Rappaport2016, Rappaport2018, Gary2017}. At some point, there was even one transit deeper than 10\% observed every 3.6 hours on average \citep{Croll2017}. As can be expected for such a system, the photosphere is highly contaminated with heavy elements \citep{Xu2016}, and it displays the typical infrared excess from the presence of a dust disk \citep{Vanderburg2015}. 

High resolution spectroscopic data also uncovered the presence of wide asymmetric circumstellar features with linewidths of $\sim 300$ km/s \citep{Xu2016}. Interestingly, the circumstellar lines have evolved from being mostly red-shifted to blue-shifted within about two years \citep{Redfield2017}. Recently, \citet{Cauley2018} proposed an eccentric misaligned precessing gas disk model to explain the evolution of the circumstellar features. While this model was successful in following the evolution of the Fe II 5316 $\AA$ region over almost a two year period, it could not provide strong insights on the physical properties of the gas (abundances, temperature and density) nor provide the expected absorption profile across the whole electromagnetic spectrum.

The goal of this paper is thus to explicitly compute the circumstellar absorption profile of WD 1145+017 at any time, from the UV to the optical, for any given geometrical/physical structure given as input. As a first step, we fully explore the configuration presented by \citet{Cauley2018} in order to test its validity, limits and finally propose modifications and avenues of research for future studies.

The observational data we used are presented in \S~\ref{sec:observations}. An updated analysis of the photospheric chemical composition of WD 1145+017 is presented in \S~\ref{sec:photosphere} while \S~\ref{sec:model} describes in detail the theoretical framework for our modeling of the precessing gas disk. Our results, including an analysis of the physical properties of the disk, its precession period, the need for additional components and the circumstellar absorption behavior during transits are presented in \S~\ref{sec:results}. Finally, we present a summary of our findings and conclusions in \S~\ref{sec:conclusion}.

\section{Observations}
\label{sec:observations}

This study uses numerous intermediate/high resolution (R = 14,000-40,000) spectra from Keck I telescope taken with the High Resolution Echelle Spectrometer (HIRESb, HIRESr) and the Echellette Spectrograph and Imager (ESI) instruments. We also use data taken with the VLT (X-SHOOTER),  HST (COS, March 28, 2016 data) as well as a light curve (taken simultaneously to Keck data) obtained from the University of Arizona's 61 inch telescope. In total, there are 17 epochs of spectroscopic observations between April 2015 and May 2018. Details concerning these observations and the data reduction procedures have already been described at length in \citet[][see also their Table 2 and 3]{Xu2019}.

\section{Photospheric Abundances}
\label{sec:photosphere}

The first detailed photospheric abundance analysis of WD 1145+017 was presented by \cite{Xu2016}. This analysis, however, assumed the atmospheric parameters determined in \citet{Vanderburg2015}, namely \Te = 15,900K and \logg = 8.0. Since this effective temperature determination was based on models that included an approximate amount of heavy elements (the circumstellar contamination of many lines was not known at the time) and assuming \logg = 8.0 \citep[parallax measurement only became available with $Gaia$ DR2 data release,][]{Gaia2018}, the atmospheric parameters and abundance analysis need to be revisited. Using the same method and grids presented in detail in \cite{Coutu2019}, we now obtain \Te = $14,500 \pm 900$ K and \logg = $8.11 \pm 0.02$. Using those new parameters, we then calculate grids of synthetic spectra for each element and determine chemical abundances using, as in \cite{Xu2016}, lines not contaminated by circumstellar absorption in HIRES data taken in March and April 2016 (2016.02.03 and 2016.03.02). While the strength and position of the circumstellar features change significantly on timescales of minutes to months \citep{Redfield2017}, we find no variation in abundances (from non-contaminated absorption lines) when determined from spectra taken at any other epochs compared to those determined from the March and April 2016 data. 

\begin{figure}[h]
    \centering
    \includegraphics[width=\columnwidth]{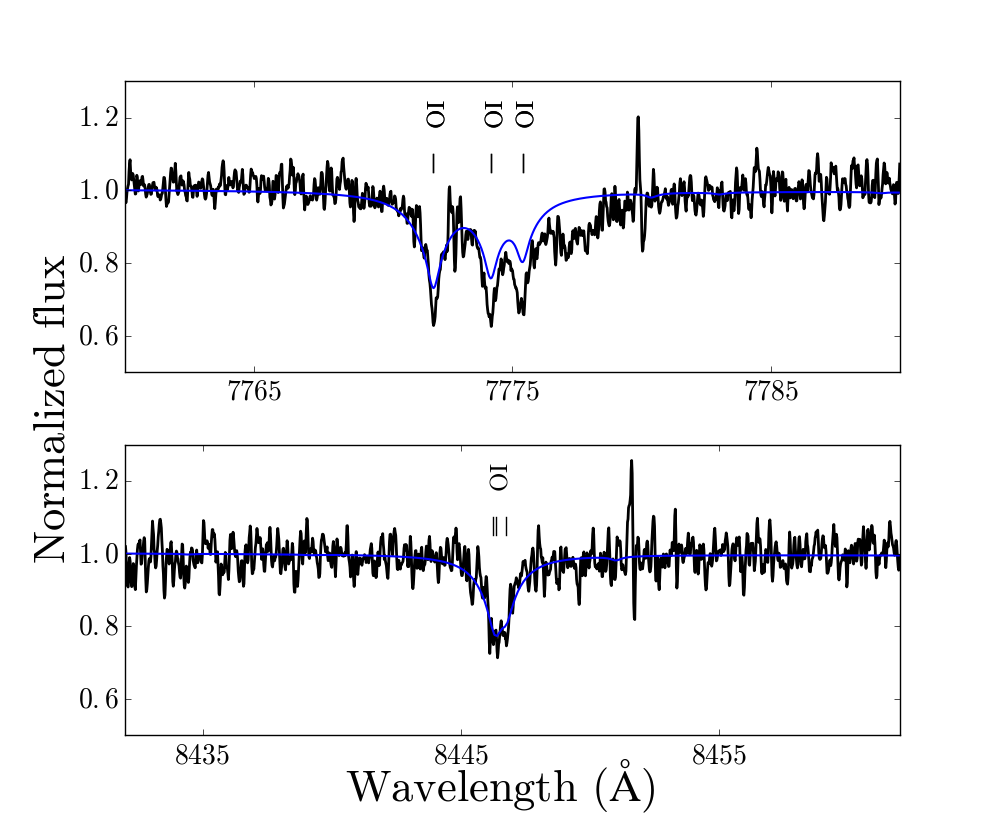}
    \caption{Oxygen absorption in the HIRESr data (2016.04.01). The top panel shows the O I triplet contaminated by circumstellar absorption. The bottom panel shows the non-contaminated O I 8446 $\AA$ line that is  used for abundance determination ($\log{\mathrm{O}/\mathrm{He}} = -5.12$, blue line)}
    \label{fig:Oxygen_newabn}
\end{figure}

\begin{figure}
    \centering
    \includegraphics[width=\linewidth]{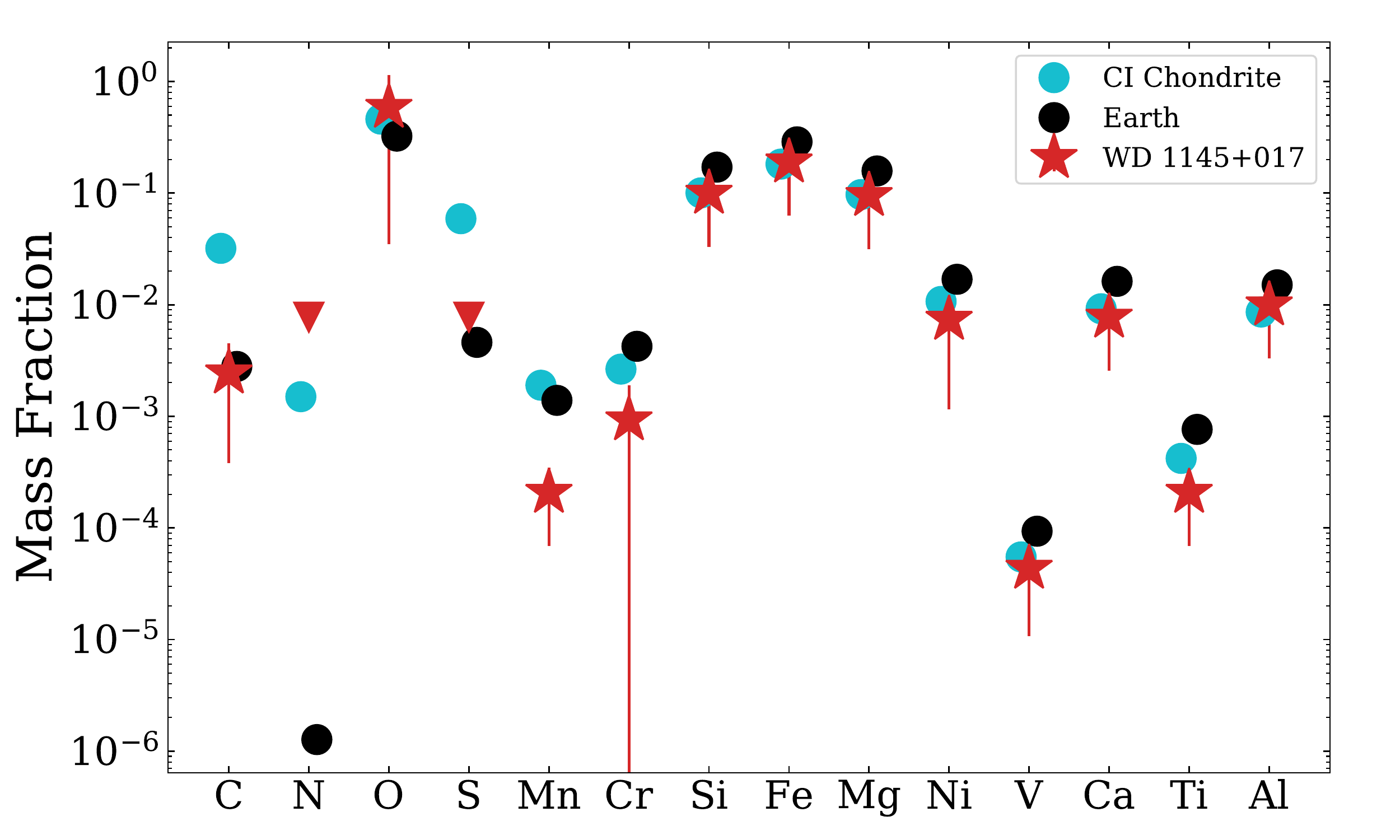}
    \caption{Mass fraction of elements accreted in the photosphere of WD 1145+017.
    }
    \label{fig:Abundance_4}
\end{figure}

\citet{Xu2016} reported conflicting results regarding oxygen, as the 2 available lines they detected in low resolution spectra indicated very different 
abundances ($\log{\mathrm{O}/\mathrm{He}} = -3.7$ for the O I 7775 $\AA$ triplet  and $\log{\mathrm{O}/\mathrm{He}} = -4.5$ for the O I 8446 $\AA$ line, see their Figure 1).
This discrepancy was concerning because it was originally thought that the high abundance from the triplet could not be explained by circumstellar absorption since these lines originate from energy levels $\sim 9$ eV above the ground level. However, there is now evidence suggesting the presence of a hot component in the system (for example, the presence of Si IV lines in the UV, see section \ref{subsec:Circular ring components}). Moreover, the new higher resolution data from HIRESr ($2016.03.02$ epoch) clearly show that the red portion of the O I triplet is contaminated by circumstellar absorption (see Figure \ref{fig:Oxygen_newabn}).  Hence, we decided to use only the O I $8446$ $\AA$ line for the photospheric abundance determination and our updated value is now $\log{\mathrm{O}/\mathrm{He}} = -5.12 \pm 0.35$. With this updated oxygen abundance measurement, the mass fraction of oxygen, which was extremely high ($\sim$ 60\%) according to \cite{Xu2016}, is now compatible with the expectation from accretion of bulk Earth-like material (see Figure \ref{fig:Abundance_4}).

\begin{figure}[h]
    \centering
    \includegraphics[width=1.05\columnwidth]{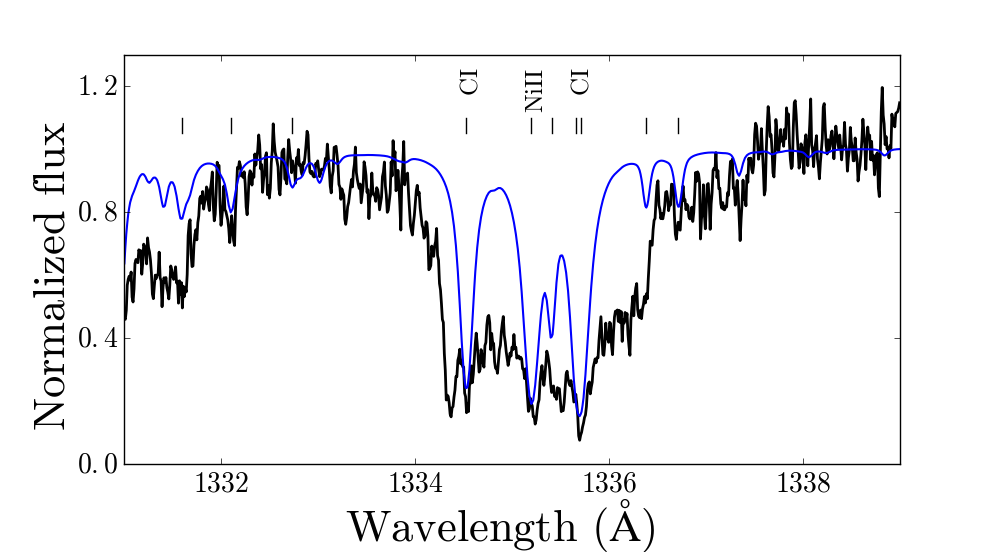}
    \caption{Possible detection of photospheric carbon (see core of strongest predicted lines) blended with circumstellar absorption features. Unidentified tickmarks indicate the position of iron lines. 
	}
    \label{fig:Carbon_43vs75}
\end{figure}

Using HST data in the UV (2016.03.28), we are also able to obtain limits for N and S. Carbon is also possibly detected (although the lines are severely contaminated by circumstellar absorption, the presence of core features with the correct radial velocity suggests that the carbon is indeed photospheric, see Figure \ref{fig:Carbon_43vs75}).

Our final photospheric abundance determinations (or limits) for each element are presented in Table \ref{tab:new_abundances}. Finally, we measure that the radial velocity from photospheric lines in the optical (too much contamination from circumstellar disk for reliable measurements in the UV ) is $ \rm 43 \pm 2 \: km/s$, which corresponds, once we remove the gravitational contribution of $\rm 35.3 \pm 0.8 \: km/s$, to a proper motion of  $\rm 8 \pm 3 \: km/s$.

\begin{table}\label{tab:new_abundances}
\begin{center}
\caption{WD 1145+017 atmospheric parameters and accretion rates}
\begin{tabular}{lllllll}
\hline \hline
\Te & 14,500 $\pm$ 900 $K$\\
\logg & 8.11 $\pm$ 0.02 \\
\hline 
Ion & log n(Z)/n(He)	& M$^{a}$	&$\dot{M}$$^{b}$ \\
	&	& (10$^{20}$ g)	& ( 10$^8$ g s$^{-1}$)\\
\hline
H			& -5.0 $\pm$ 0.20 	& ... & ... & \\
C		 	& $\sim -7.5$ 	    & 1.9  & 0.0837  \\
N           &  $<-7.0$          & $<$6.9  & $<$0.339  \\
O			& -5.12 $\pm$ 0.35  & 600   & 29.9  \\
Mg		    & -5.91 $\pm$ 0.20  & 146   & 7.19   \\
Al 		    & -6.89 $\pm$ 0.20  & 17  & 0.875  \\
Si 	    	& -5.89 $\pm$ 0.20  & 179   & 8.97   \\
S 	    	& $<-7.0$           &$<$16  & $<$0.963  \\
Ca	        & -7.0 $\pm$ 0.20   & 20  & 0.151  \\
Ti 		    & -8.57 $\pm$ 0.20  & 0.64 & 0.0528  \\
V:		    & -9.25:$\pm$ 0.25  & 0.14 & 0.0118  \\
Cr 	        & -7.92 $\pm$ 0.40  & 3.1  & 0.248  \\
Mn		    & -8.57 $\pm$ 0.20  & 0.73 & 0.0586  \\
Fe			& -5.61 $\pm$ 0.20  & 680   & 52.4  \\
Ni			& -7.02 $\pm$ 0.30  & 28  & 0.207  \\
total   	&   ...	            & 1497 & 100 &  \\
\hline
\end{tabular}
\end{center}
{\bf Notes.} \\
$^{a}$ Current mass in the white dwarf's convection zone \citep{Dufour2017}. \\
$^{b}$ Accretion rate assuming a steady state. \\
\end{table}

%%%%%%%%%%%%%%%%%%%%%%%%%%%%%%%%%%%%%%%%%%%%%%%%%%%%%%%%%%%%%%%%%%%%%%%%%%%%%%%%%%%%%%%%%%%%%%%%%%%%%%%%%%%%%%%%%%%%%%%%%%%%%%%%%%%%%%%%%%%%%%

\section{Gas Disk Model}
\label{sec:model}

\subsection{Theoretical Framework}
\label{sec:theoretical}

In order to account for the varying asymmetrical circumstellar absorption features seen in the spectra of WD 1145+017, a simple elliptical precessing gas disk model similar to that proposed by \cite{Cauley2018} is developed. Here, we aim to explicitly compute all circumstellar absorption, from UV to optical at various times in the cycle assuming the same geometric configuration, but based on detailed opacity calculations for the physical conditions present in the disk.
The disk is presented edge-on with a non-negligible width covering about half of the star surface. Note that, as noted by \cite{Cauley2018}, this configuration is probably not the only one able to reproduce the circumstellar absorption features as the reality may certainly be more complex. However, by exploring in detail the successes and failures of this simple configuration, we can get a much deeper physical insight that will be valuable for future studies of this system. The remainder of this section provides details on the opacity calculations, the gas disk construction and free parameters that are used to adjust it to match the various epochs of observation obtained in the last 4 years. 

\subsection{Disk Configuration}
\label{sec:Geometry}

\begin{figure}[h]
    \includegraphics[width=\columnwidth]{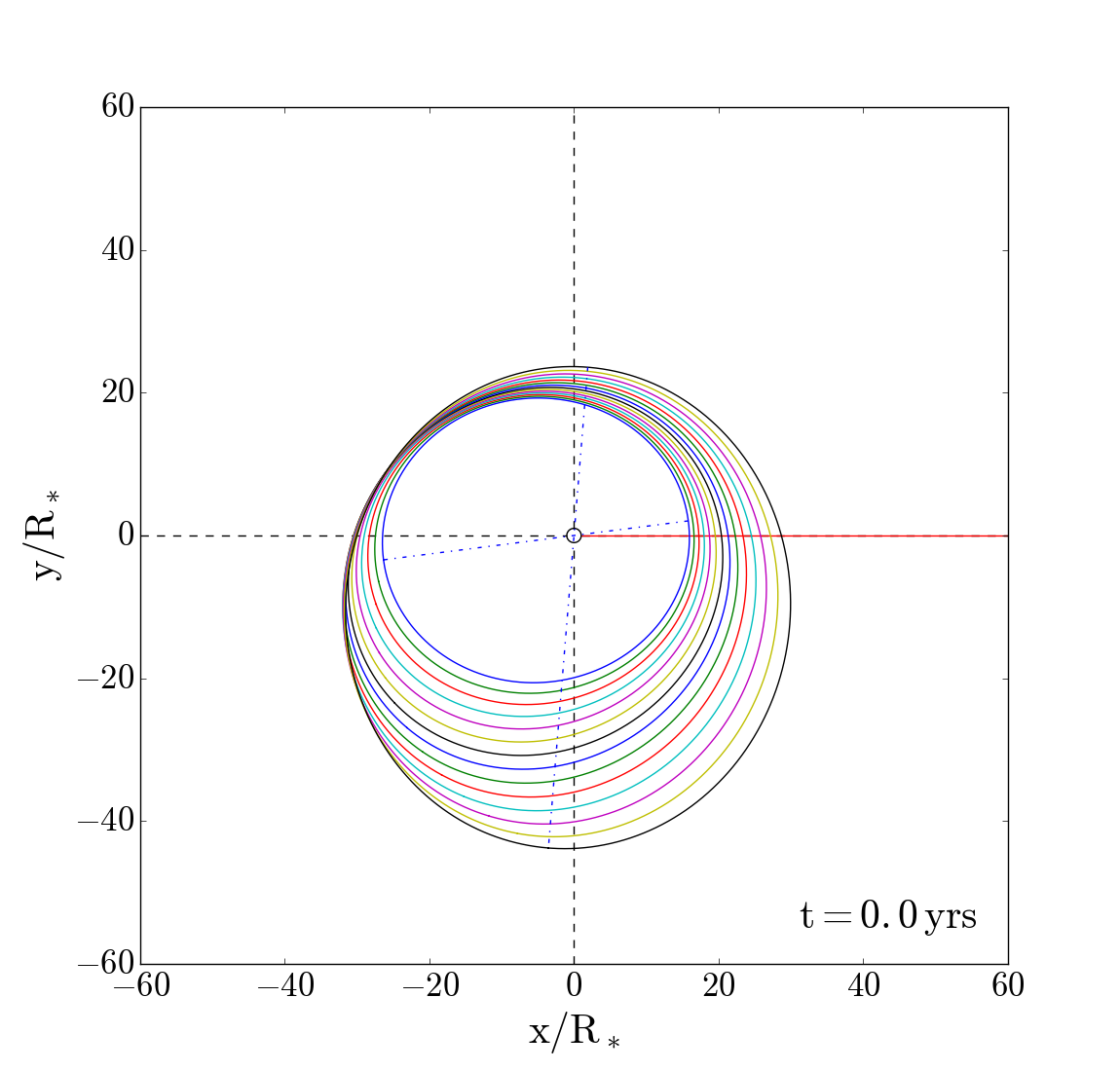}
    \caption{Configuration of the 14 eccentric gas rings. The dashed blue lines represent the apsidal line of the innermost and outermost rings, the red line is the line of sight and the black circle is the position of the star.  
	}
    \label{fig:config_geo_t0}
\end{figure}

Following \cite{Cauley2018}, the disk is constructed with 14 eccentric misaligned rings in the $xy$ plan. There are 3 parameters governing the position of the rings and they each vary linearly between their value for the innermost and outermost ring. The perihelion distances vary between $15.93 \, R_{*}$ and $23.64 \, R_{*}$, the eccentricities vary between 0.25 and 0.30 and finally, we assume a $78^{\circ}$ shift between the apsidal lines of the innermost and outermost rings. Each ring has a radial width of $0.5 \, R_*$ with $R_*=0.0118 \, R_{\odot}$, as determined from our updated stellar parameters. The 14 rings are confocal and the common focus is where the star is positioned.  An example of the configuration at an arbitrary moment in the precession cycle is 
shown in Figure \ref{fig:config_geo_t0}. The model disk is also precessing. The precession period is equal for the 14 rings, so the rotation of the configuration is solid. The absorption from this disk can be calculated at any time $t$ (or angle of rotation of the rings) to predict the shape of the circumstellar features (see section \ref{subsec:Precession period}).

\subsection{Opacity Calculation}
\label{sec:opacity}

The disk is positioned edge-on in our line of sight. To compute the total circumstellar absorption, we divide the surface of the star that is covered by the disk in a 20$\times$20 grid in the $yz$ plan. Each grid box is composed of 14 layers representing the 14 gas rings, each with an opacity specifically computed from the physical parameters at that point in the disk. We compute the radiation transmitted through the disk at each line of sight by attenuating the specific intensity from the star at that angle with an exponential $e^{-\sum_{\rm ring}^{} \kappa_{\nu} d \rm m}$, where $\kappa_{\nu}$ is the opacity for the column mass $dm$ of the ring (Doppler shifted appropriately using the velocity profile, see below) and the sum is done over the 14 ring layers. We thus do not consider emission or multiple scatterings through the rings in this simple model. The spectrum is then obtained, as usual, by integrating the specific intensities over the surface of the star (limb darkening is thus automatically taken into account in the procedure).

To compute the opacities, we need to attribute physical parameters everywhere in the disk (i.e we need the temperature, the mass density and the chemical composition to describe the gas in each cell). We first assume that the chemical composition of the disk is the same as that of the photosphere, an assumption that appears to be excellent to the first order (see section \ref{subsec:Density and Temperature}). We exclude, however, hydrogen and helium since we see no circumstellar absorption from these elements (moreover, they are not expected to be a significant part of a tidally disrupted asteroid).

For the values of the mass density, we use vertical and radial structures for the disk similar to those proposed by \cite{Cauley2018} assuming the disk temperature approximation described in \cite{Melis2010}. The density structure is computed from the input value of the density of the innermost ring in the middle plane of the disk ($z=0$), $\rho_0(r_{in})$, which is a free parameter of the model. The middle plane values for the following rings are given by 

\begin{equation}
	\rho_0(r)=\rho_0(r_{in}) \times \left (\frac{r_{in}}{r} \right ) ^2
  \label{eq:density_rprofile}
\end{equation}
where  $r$ is perihelion distance of the rings and $r_{in}$ represents the value for this innermost ring. From this, we compute the vertical scale of each ring using

\begin{equation}
	\rho(z)=\rho_0 \; e^{-z^2/H^2}
  \label{eq:density_zprofile}
\end{equation}
where $\rho_0$ is density at $z=0$ for each ring and $H$ is the scale height given by \citep{Melis2010},

\begin{equation}
	H=\left ( \frac{2 \, k \, T_{gas} \, D^3}{G \, M_* \, \mu} \right ) ^{1/2} ,
  \label{eq:scale_height}
\end{equation}
where $T_{gas}$ is the gas temperature, $D$ the distance from the star, $M_*$ the mass of the star ($0.656 \: M_{\odot}$) and  $\mu=2.4 \times 10^{-23}$ g. The density profile is symmetric around the $z=0$ plane and is also equal around a ring. 

For simplicity, we first use a constant temperature throughout the disk. While this approximation allows a fair representation, at certain epochs, for most circumstellar absorption features in the optical, this simple model is clearly insufficient to reproduce all features, indicating that a more complex temperature structure is needed. We thus also experiment with some temperature gradients in the gas (see section \ref{subsec:gradient}).

Using the physical structure for the disk described above (temperature, density and abundances), we compute the opacity of the rings in each grid box using the public atmospheric code package \textsc{tlusty/synspec}\footnote{http://nova.astro.umd.edu/index.html}. The synthetic spectrum code \textsc{synspec} has a special mode called "iron curtain", which takes the temperature, electronic density and chemical composition as input to compute the opacity of an uniform slab of gas with those parameters. The electronic density, an a priori unknown quantity, is obtained by first running the atmospheric code \textsc{tlusty} with a one layer input having the temperature, mass density and abundances fixed to the desired values.

\begin{figure}[h]
    \includegraphics[width=\columnwidth]{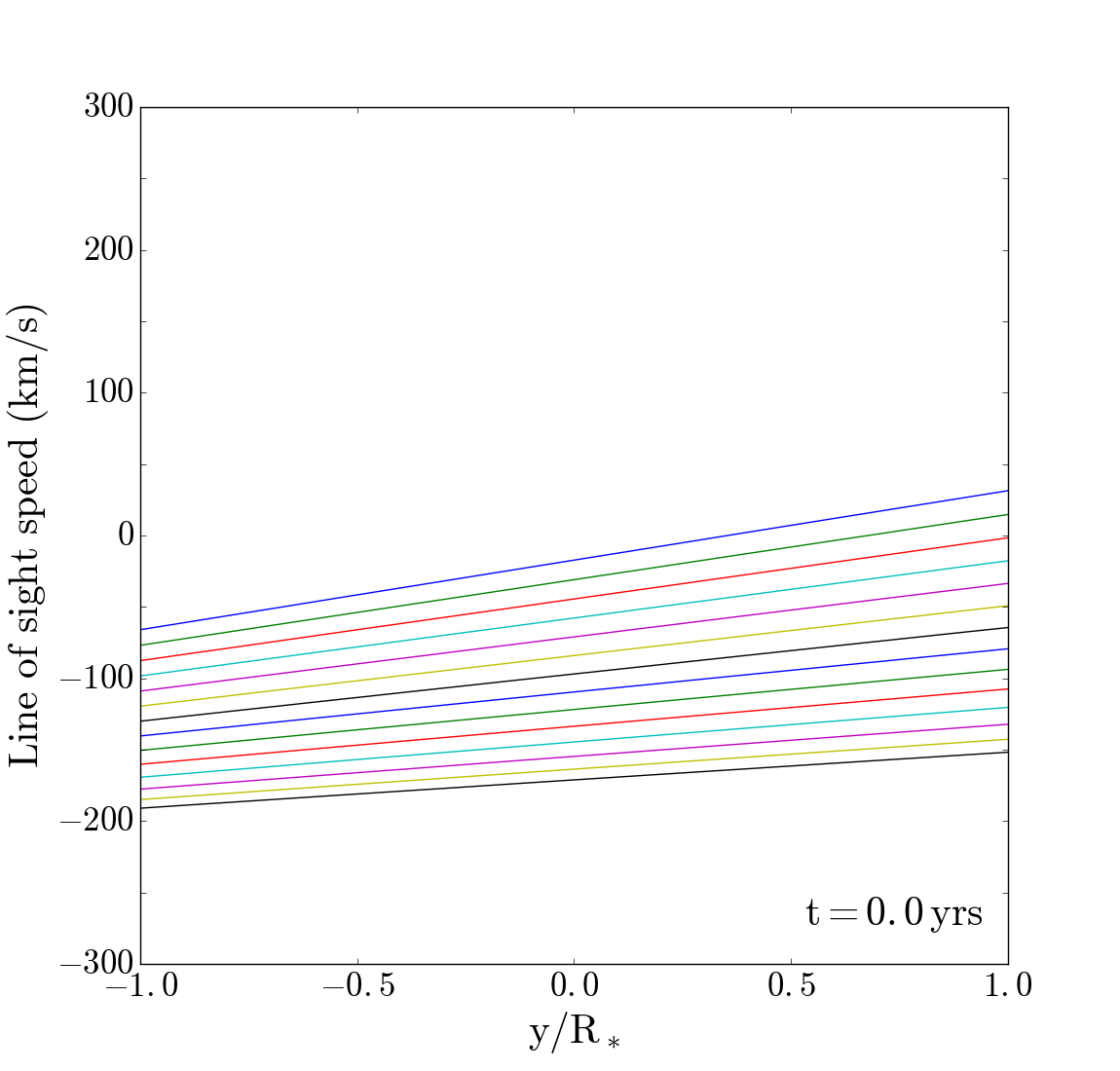}
    \caption{Radial velocity profile of each ring across the line of sight for the $t=0$ yrs configuration. The colors for each ring matches the colors used in Figure \ref{fig:config_geo_t0}. 
	}
    \label{fig:velocity_profile_t0}
\end{figure}

Once the opacity in each grid cell has been calculated, Doppler velocity shift due to the revolution of the gas in the rings is applied. The line of sight velocity is simply the keplerien orbital velocity ($x$ component)

\begin{equation}
	v=\sqrt{G M_* \left ( \frac{2}{r} - \frac{1}{a} \right )} .
  \label{eq:kepler_velocity}
\end{equation}
where  $r$ is the distance from the star, and $a$ the semi-major axis of the ellipse.  

The line of sight velocity profile for the geometric configuration displayed earlier is presented in Figure \ref{fig:velocity_profile_t0}. We can then see that for this particular configuration, the profile is almost entirely red-shifted. As the disk precesses, different ranges of line of sight velocities can be obtained, producing, as observed for WD 1145+017, circumstellar feature shifts from $\sim$+200 km/s to -200 km/s.

\begin{figure}[h]
    \includegraphics[width=\columnwidth]{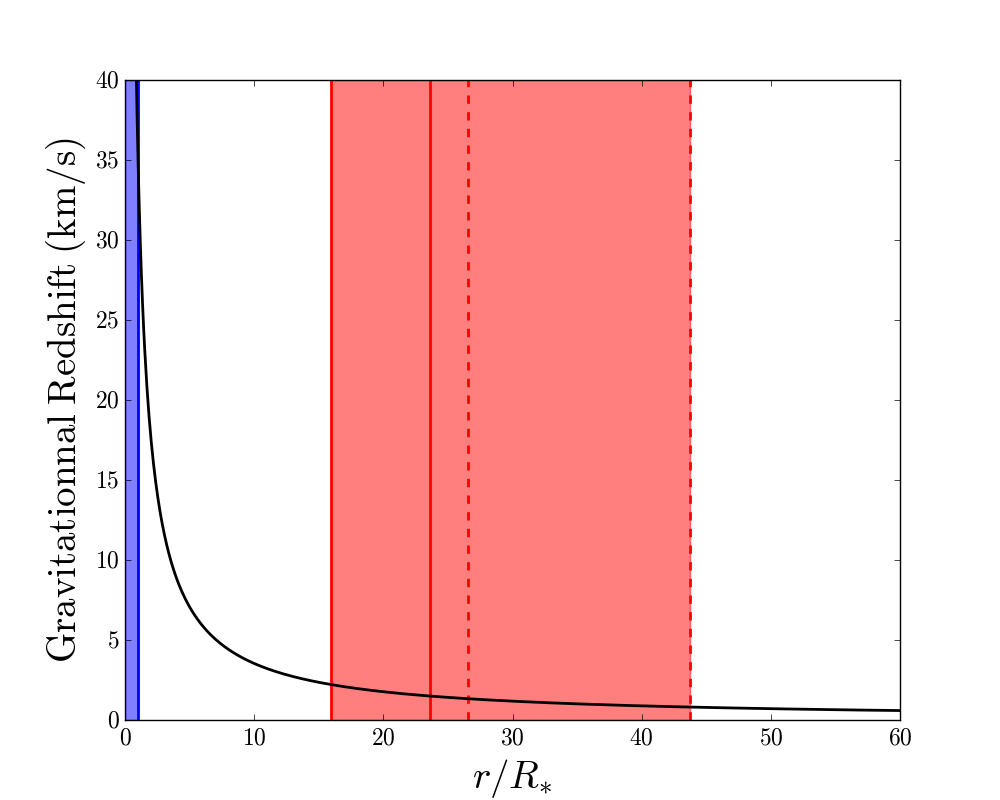}
    \caption{Gravitational redshift as a function of distance from WD 1145+017 (black curve). The blue area represents the position of the star and the red area the position of the disk. The full red lines show the position of the perihelion for the innermost and outermost rings, and the dashed red lines show the aphelion distances. 
	}
    \label{fig:redshift_grav}
\end{figure}

Finally, although only a small component, we also include the gravitational redshift, $v_{grav}=\frac {G M_*} {c \, r}$ (see Figure \ref{fig:redshift_grav}). This additional shift slightly changes the shape of the circumstellar features as the velocity shift varies along a ring and is different for each ring.

\begin{figure}
    \centering{
    \includegraphics[width=\columnwidth]{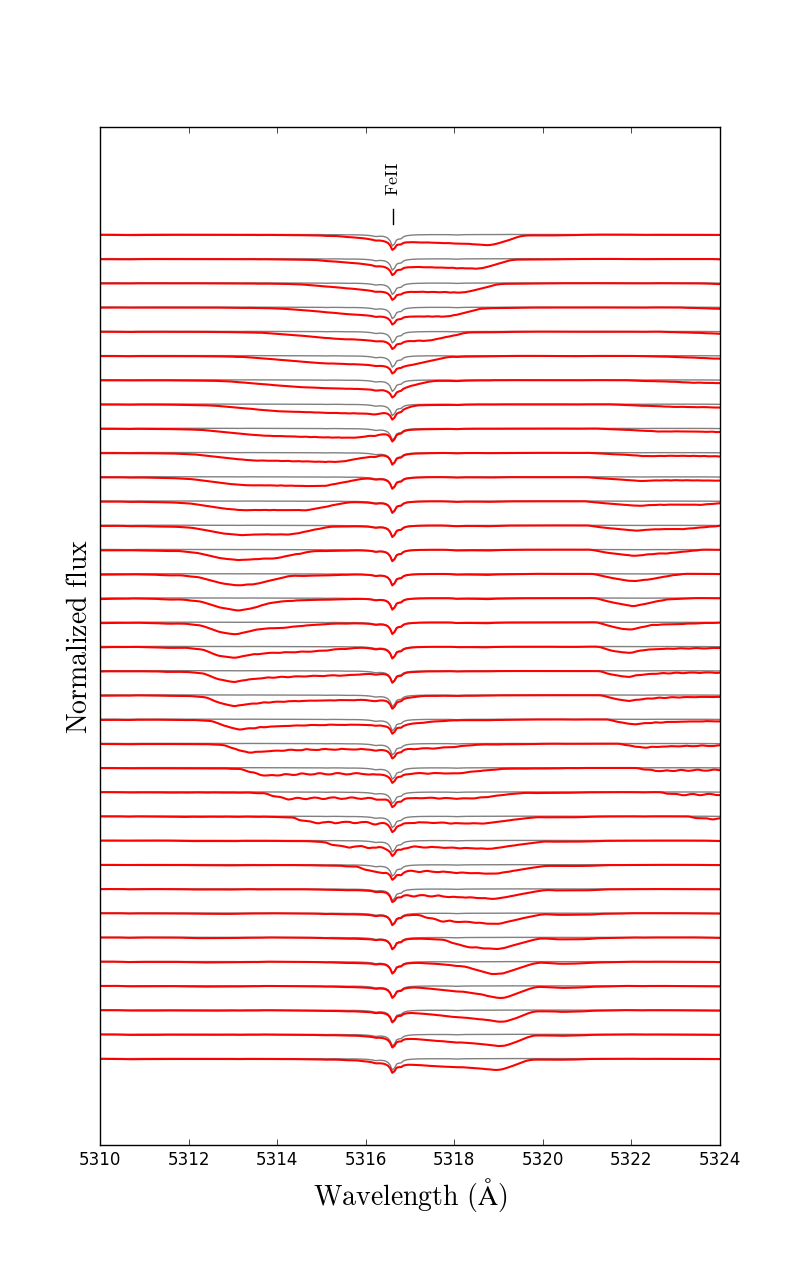}
    \caption{ Precession cycle in steps of about 10$^{\circ}$ for the $5316.6$ $\AA$ FeII line. In gray is the photospheric model and in red is the precessing disk model.
	}}
    \label{fig:article_precessioncycle}
\end{figure}

Figure \ref{fig:article_precessioncycle} shows an example of the resulting circumstellar absorption for the 5316 $\AA$ FeII line during the entire precession cycle of the disk. In the next sections, we compare in detail the predictions of this simple model with absorption features observed at different epochs, from both optical and UV data.

\section{Results}
\label{sec:results}

\subsection{Density and Temperature}
\label{subsec:Density and Temperature}

\begin{figure*}
    \centering{
    \includegraphics[width=\textwidth]{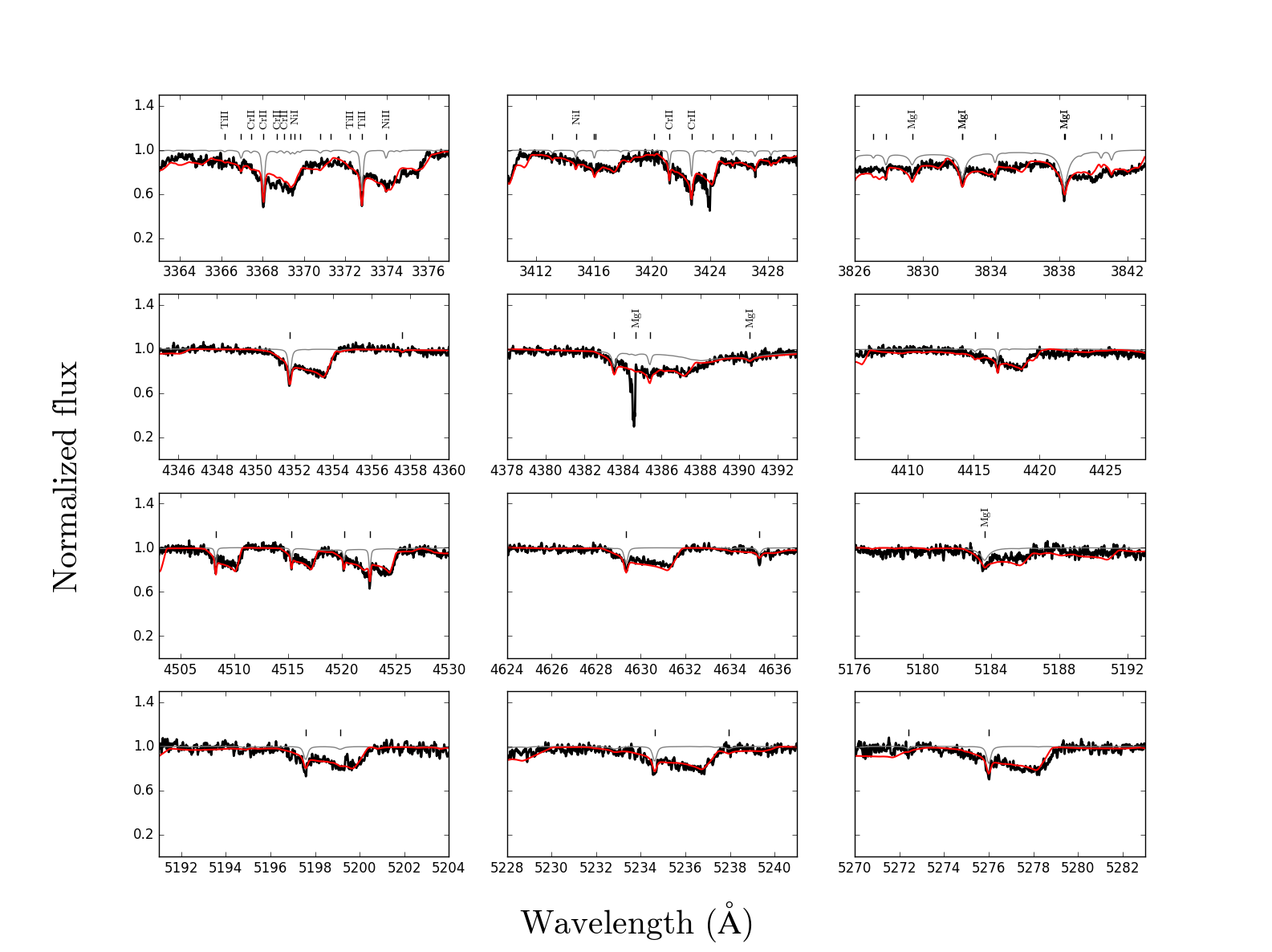}
    \caption{Selected regions from the HIRESb 2016.04.01 data (black). The modeled circumstellar features (red) are shown for a central density of $6 \times 10^{-6}$ $\rm g/cm^3$ and a temperature of $6000$ $\rm K$. The gray line represents the photospheric contribution alone. The unidentified lines are iron.
	}}
    \label{fig:article_raiesbien6000}
\end{figure*}

We first determine the combination of free parameters that best represent the shape and depth of the circumstellar features. We start with the HIRESb optical spectra for the 2016.04.01 epoch which presents almost entirely red-shifted features. We first find the time (angle) in the precession cycle that best reproduces the overall shape of the absorption features. This time is henceforth used as the zero point of the precession period. Once the disk is well positioned, we try different combinations of central density and gas temperature in order to reproduce the depth of all circumstellar features simultaneously.  We test wide ranges from 3000 K to 30,000 K (in step of 1000 K) and $1 \times 10^{-7}$ $\rm g/cm^3$ to $5 \times 10^{-5}$ $\rm g/cm^3$. We find that the combination that best reproduces the shape of a fair amount of the circumstellar features is a central density of $(6.0 \pm 1.0)  \times 10^{-6}$ $\rm g/cm^3$ and a temperature of $6000 \pm 1000 $ $\rm K$, which we will use in what follows (the quoted uncertainties are conservative values based on clearly inferior fit using adjacent grid points in our parameter space). We note that the total integrated mass for the eccentric rings assuming this structure is $2.1\times 10^{16}$ g (somewhere between the mass of Uranus and Neptune's rings, or about a millionth of the mass of Saturn's rings), a value that should only be considered a rough order of magnitude estimate, given all the approximations involved. Nevertheless, this is significantly smaller than the total amount of material present in the star's photosphere (see Table \ref{tab:new_abundances}). Also, the lifetime of the gas, given by the estimated mass of the gas disk divided by the total accretion rate (derived assuming steady state), would be much less than a year, indicating that the gas must be replenished completely on a very short timescale \citep[see also][for a similar discussion]{Xu2016}. However, since it has been recently proposed that convective overshooting may also have a significant effect on the diffusion coefficients and mass of the mixed region in helium-rich white dwarfs \citet{Cunningham2019}, it is probably best not to acquiesce literally yet to interpretations based on 1D convection zone models. 

In Figure \ref{fig:article_raiesbien6000}, we show examples of lines that are reproduced very well with this simple model. It is interesting to note that the depths for several elements (Mg, Ca, Cr, Ti, Fe, Ni) are all reproduced simultaneously with this model, indicating that the assumed input abundances, which are the the ones found from the photospheric analysis, are an excellent first order approximation.

\begin{figure*}
    \centering{
    \includegraphics[width=0.87\textwidth]{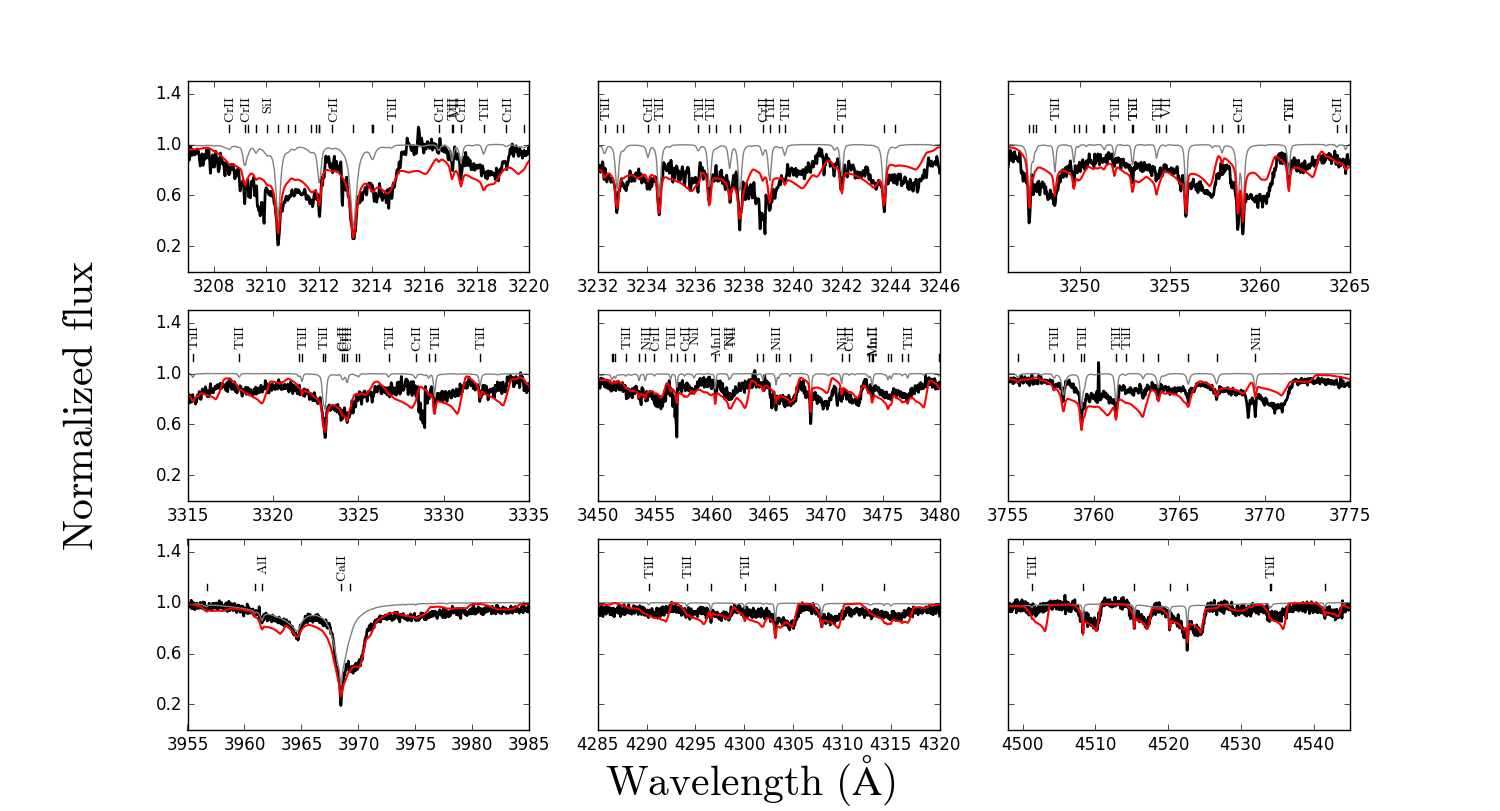}
    \caption{Selected regions in the HIRESb 2016.04.01 spectrum badly reproduced by the constant temperature model. The red and grey lines are the same models presented in Figure \ref{fig:article_raiesbien6000}.
	}}
    \label{fig:article_raiesmauvaisesdeuxtemps}
\end{figure*}

There are, however, regions where the model does not match as well, or predicts lines that are not observed (see Fig \ref{fig:article_raiesmauvaisesdeuxtemps}). While these lines appear to require a different gas temperature to be reproduced, we found no constant temperature model that is satisfactory.

\begin{figure*}
    \centering{
    \includegraphics[width=0.9\textwidth]{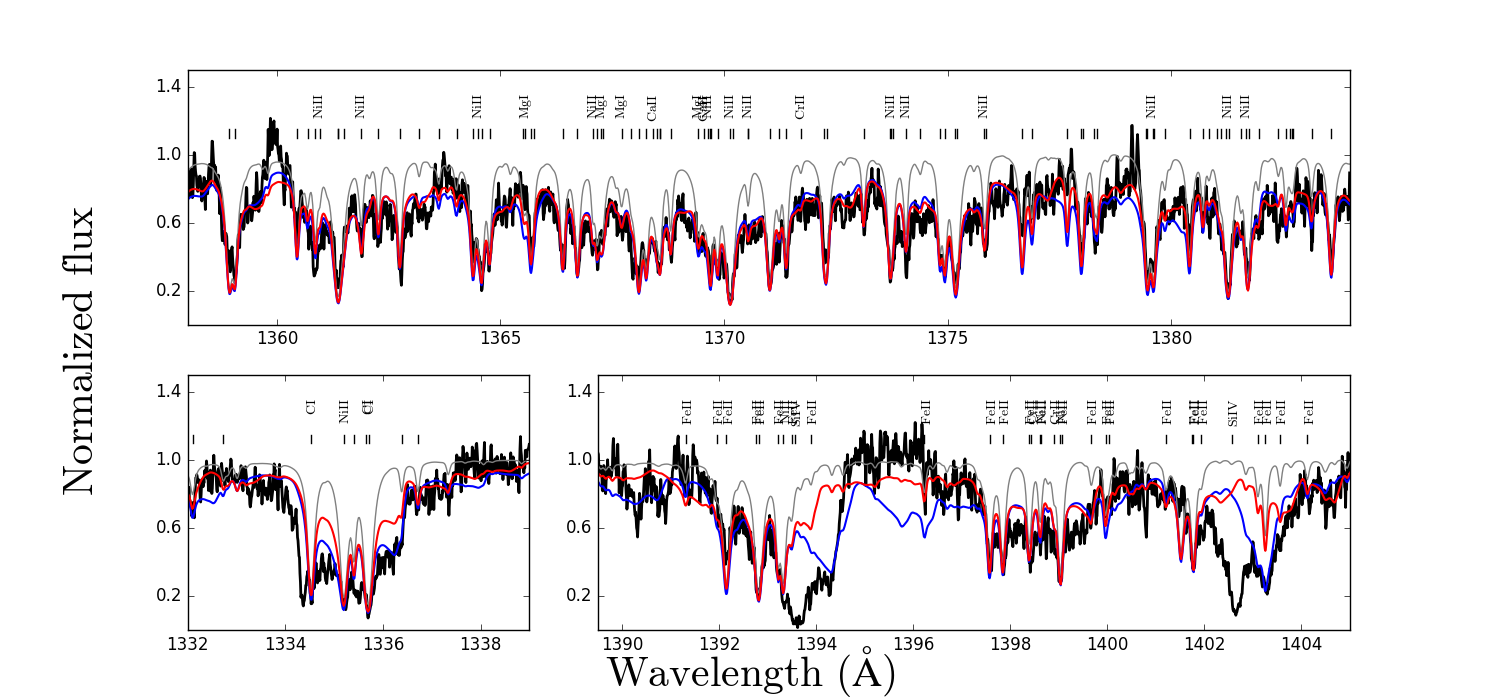}
    \caption{Selected regions from the 2016.03.28 HST spectrum. The red and grey lines are the same models presented in Figure \ref{fig:article_raiesbien6000} and the blue line shows the combined model of the star photosphere and the disk for a temperature of $13,000$ $\rm K$. The unidentified lines are iron. 
	}}
    \label{fig:article_raiesbien13000}
\end{figure*}

We can also compare the prediction from our model with data taken in the UV with HST only a few days before the HIRESb data. Since the number of transitions is much greater in the UV, the circumstellar and stellar features cannot be isolated and they practically form a superposed continuum of lines. Nevertheless, Figure \ref{fig:article_raiesbien13000} shows that the $6000$ $\rm K$ model reproduces most of the absorption quite successfully with the exception of a few notable lines arising from higher energy levels. In particular, we note the presence of two strong Si IV lines ($\sim 1394$ and $1402\AA$), indicating that a much higher temperature is needed \citep[note that similar circumstellar Si IV lines were also detected in two hotter polluted white dwafs, PG 0843+516 and SDSS 1228+1040, see][]{Gaensicke2012}. Increasing the temperature to $13,000$ $\rm K$ can produce such strong lines (it also increases the depth of the circumstellar carbon component) without causing too many changes in the parts that were previously well reproduced with the $6000$ $\rm K$ model (see blue line in Figure \ref{fig:article_raiesbien13000}). 

Although the hotter disk model accounts for these particular lines, it also predicts features where there are none, for example the Fe line we see around $1396$ $\AA$. This indicates that our constant temperature approximation is not sufficient to simultaneously explain all the circumstellar features.

We also note the presence of unaccounted for symmetric components not red-shifted (the strong lines near $\sim 1394$ and $1402\AA$ that are not reproduced by either temperature models). These features most probably originate from a circular, or very low eccentricity, gas ring situated further out than the eccentric disk. Our analysis of these "zero shift/circular components" is presented in section \ref{subsec:Circular ring components}.

\subsection{Disk Structure with Temperature Gradient}
\label{subsec:gradient}

\begin{figure*}
    \centering{
    \includegraphics[width=0.87\textwidth]{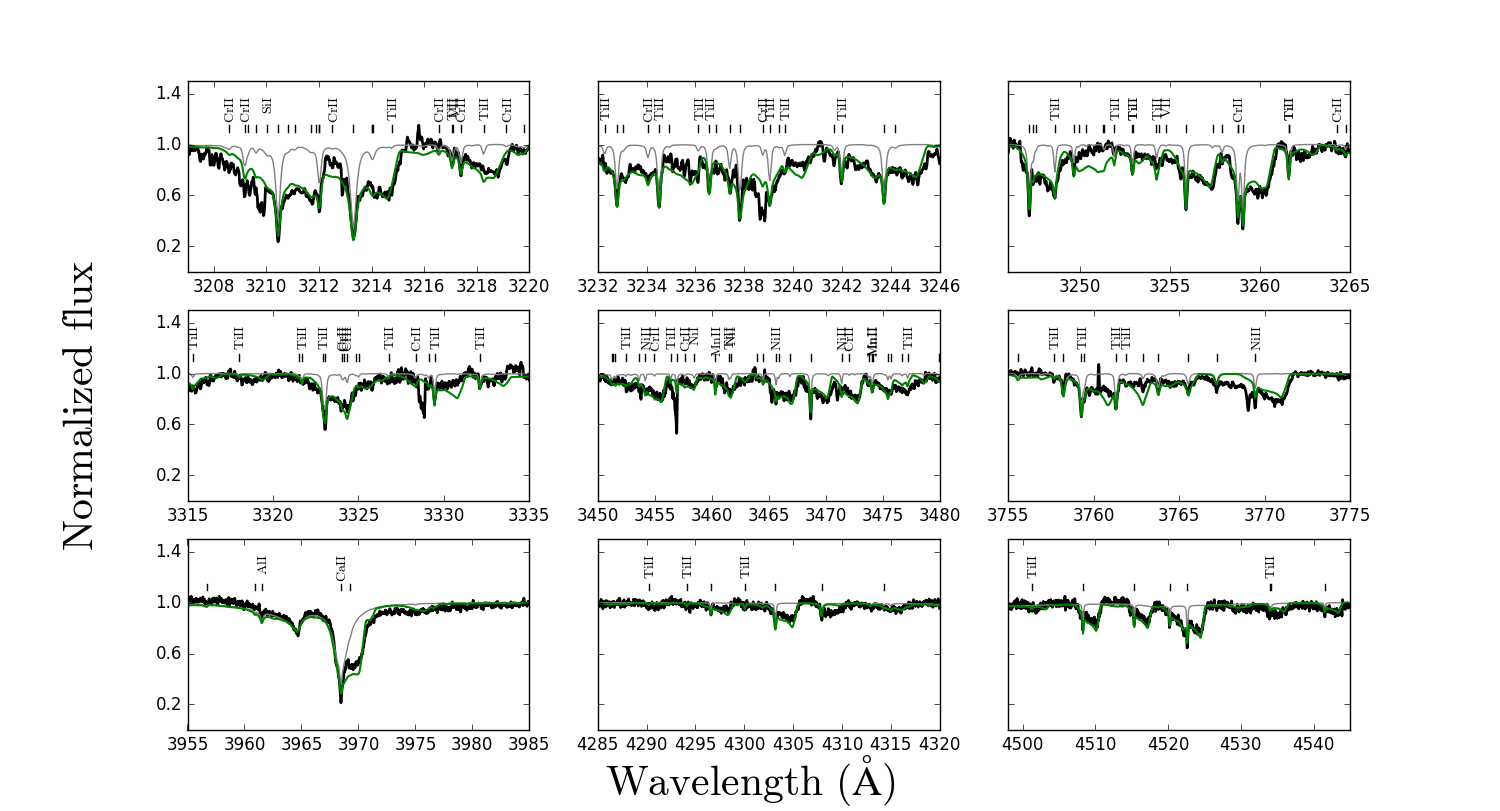}
    \caption{ Same regions of the HIRESb 2016.04.01 spectrum as presented in Figure \ref{fig:article_raiesmauvaisesdeuxtemps} with our model including a vertical temperature structure (green). The unidentified lines are iron. 
	}}
    \label{fig:article_raiesmauvaisesstructuretemp}
\end{figure*}

\begin{figure*}
    \centering{
    \includegraphics[width=0.9\textwidth]{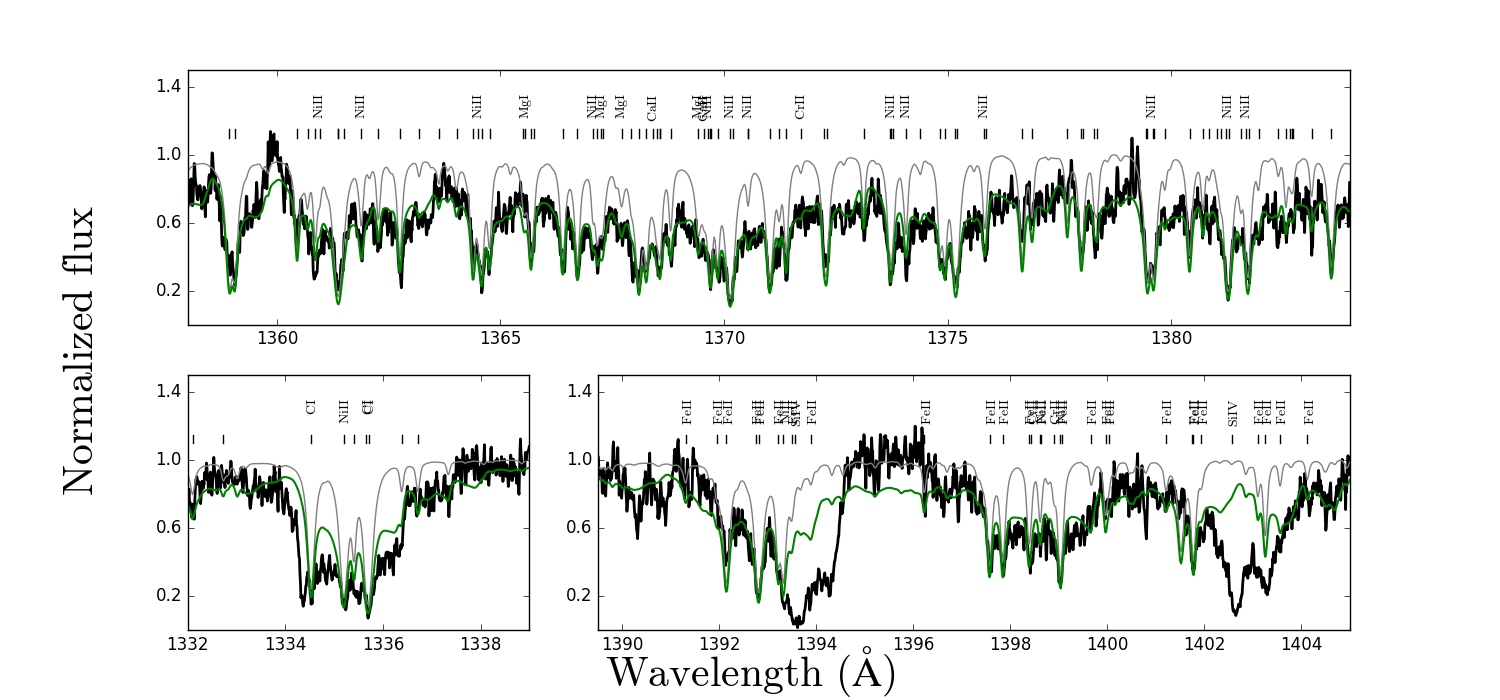}
    \caption{Same regions of the 2016.03.28 HST spectra as presented in Figure \ref{fig:article_raiesbien13000} with the temperature structure model in green.
	}}
    \label{fig:figure9_struc_ld}
\end{figure*}

The computation structure of our model allows us to look at the opacity in each grid box in order to track where particular lines are formed and thus assess what the physical parameters needed to form them are. We find that the denser central parts of the disk need to be hotter to form lines from higher energy levels, and the outer parts of the disk need to be cooler to create the many less energetic lines. We thus use this information to test different simple temperature structures that have hotter central regions, and cooler outer regions. 

\begin{figure*}
    \centering{
    \includegraphics[width=\textwidth]{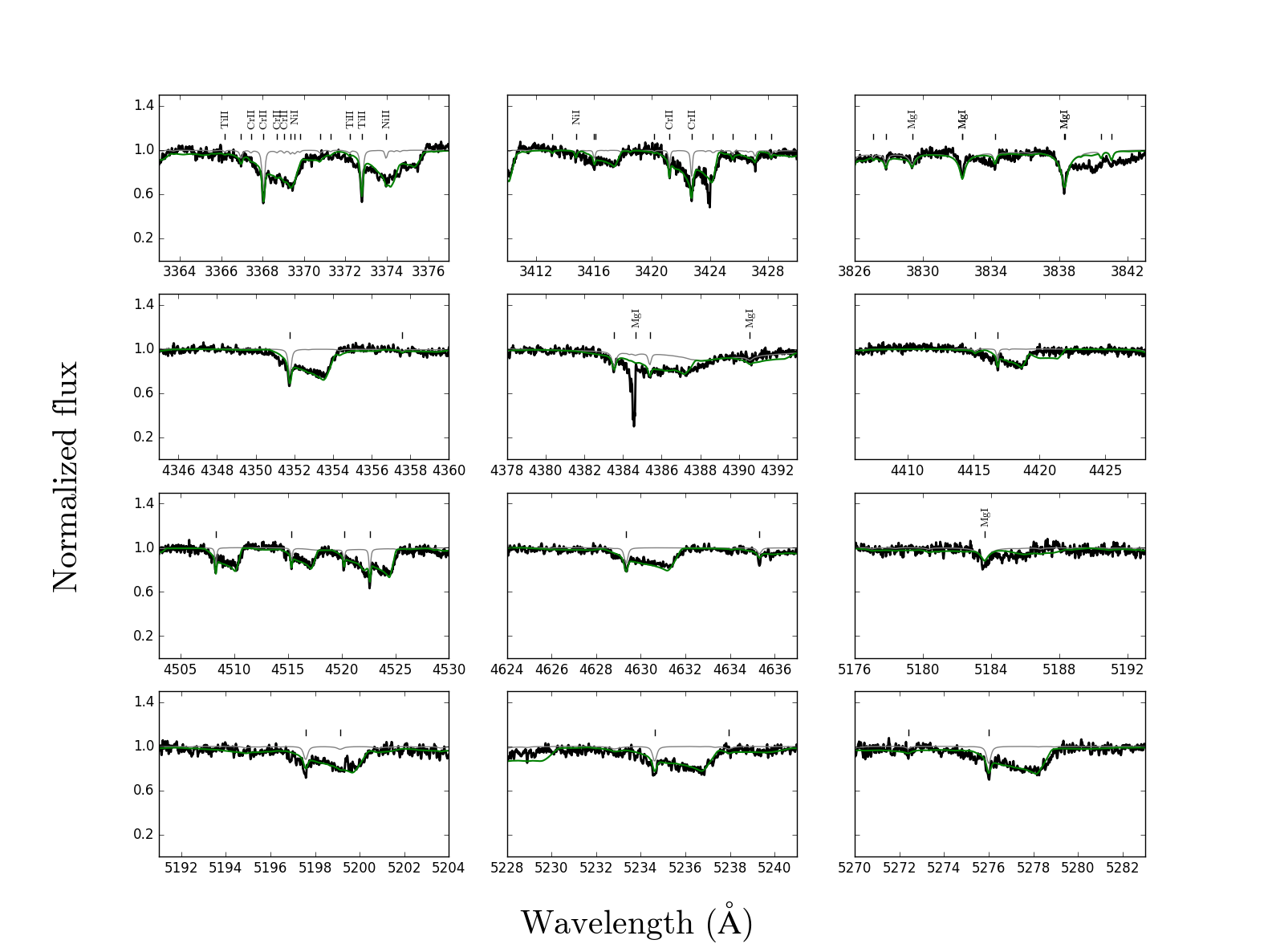}
    \caption{Same regions of the 2016.04.01 HIRESb data epoch presented in Figure \ref{fig:article_raiesbien6000}, with our model including a vertical temperature structure (green). 
	}}
    \label{fig:figure8_struc_20160401}
\end{figure*}

As described in section \ref{sec:Geometry}, the vertical width of the disk is divided in 20 boxes that have a density structure symmetric around the central plane. We explore temperature structures constructed the same way, that is they are also symmetric around the central plane, but for the sake of simplicity we keep this vertical structure constant throughout the 14 rings (thus 10 vertical layers, from the central to the outer temperature values). We also simply assume, for this little experiment, a linear variation of the temperature. We attribute a $13,000$ $\rm K$ value for the central layer and the 9 others are scaled linearly with a step of $1100$ $\rm K$, bringing the outer temperature to $3100$ $\rm K$. These parameters were obtained by testing different central temperatures and steps and comparing the resulting models with certain features in the data that are affected by these changes. The central density is kept at $6 \times 10^{-6}$ $\rm g/cm^3$, and since the vertical density structure of each ring depends on the gas temperature (see scale height Eq. \ref{eq:scale_height}), we use the mean value of the structure, which in this case is $8050$ $\rm K$. We note that this simple temperature structure is not physical, and that a detailed analysis using a temperature structure obtained from first principle, which is beyond the scope of this paper, should eventually be performed. Nevertheless, this experiment is very useful to gain some valuable insights on the physical conditions present in the disk. 

Results from this non-constant temperature model are shown in Figure \ref{fig:article_raiesmauvaisesstructuretemp} and \ref{fig:figure9_struc_ld} for the same regions presented Figure \ref{fig:article_raiesmauvaisesdeuxtemps} and \ref{fig:article_raiesbien13000}. Although the agreement is much better than the constant temperature models, there are still a few problems. In particular, the lines from higher lower energy level displayed in Figure \ref{fig:article_raiesbien13000} are still not deep enough even with the much hotter central regions of the disk. As for the regions that were well reproduced with the $6000$ $\rm K$ model in Figure \ref{fig:article_raiesbien6000}, the change in temperature structure has little impact on those lines, with the exception of a few (for example, magnesium near 3840 \AA, see Figure \ref{fig:figure8_struc_20160401}). This simple model also produces an O I absorption feature similar to what is observed in the $\sim$ 7775 \AA~  triplet (see Figure \ref{fig:Oxygen_newabn}), although it is predicted too deep.

These shortcomings are not surprising given that the structure we implemented is very simple and probably not realistic. The vertical temperature scale is probably not linear and there must also be a radial temperature scale. We also do not take into account that the rings are eccentric, meaning there are parts of the same ring that are closer to the star, and thus probably hotter, than others. It is also possible that the chemical composition of the disk is not exactly the same as that determined in the photosphere, thus affecting the relative depths of some circumstellar features from different elements. Nevertheless, considering all those approximations, this simple model does a fairly good job at reproducing the majority of the circumstellar absorption features from the UV to the optical, while providing, at the same time, some interesting insights to the real physical structure of the disk.

\subsection{Zero Velocity Absorption Component}
\label{subsec:Circular ring components}

\begin{figure}
    \centering{
    \includegraphics[width=\columnwidth]{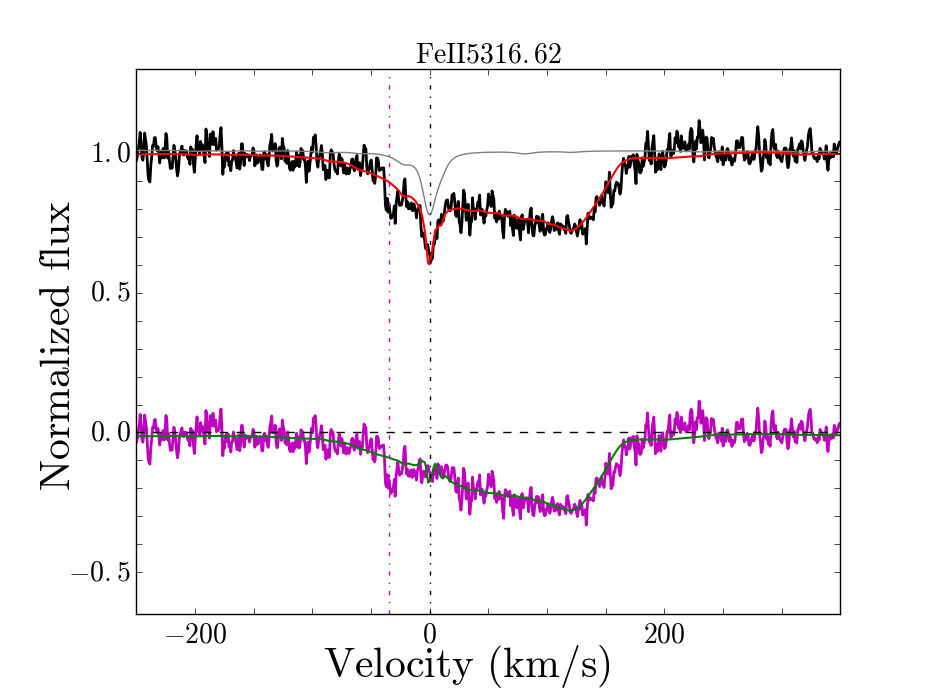}
    \caption{Example of a "zero velocity" component for the $5316.6$ $\AA$ Fe II line. Top: photospheric (gray) and gas disk model (red). Bottom: gas disk model (green) and data (magenta) from which we subtracted the photospheric model in order to better display the additional component (magenta dotted line at minus the gravitational redshift). 
	}}
    \label{fig:article_5316tick}
\end{figure}

We briefly mentioned, in the previous section, the presence of components with practically zero velocity shift that were contributing significantly to the Si IV profiles. Similar stationary components, although not as strong as the silicon ones, can also be observed in the circumstellar absorption profiles of many elements in the optical (see Figure \ref{fig:article_5316tick} for an example). Such absorption features cannot be reproduced using our eccentric disk model. These features are relatively narrow, symmetric and blue-shifted by almost exactly the value, within the uncertainties, of the gravitational redshift, indicating that the source of this absorption is located at a distance where the redshift is $\sim 1$ km/s at most. These features can thus be used to confirm (or determine independently) the mass of the white dwarf. These blue-shifted (relative to the photosphere) features are present at all the epochs and appear to be stable. A similar feature was also detected for the H and K Ca lines in the spectra of the white dwarf WD 1124-293 by \cite{Debes2012}, which they interpreted as evidence for circumstellar gas. 

The value of the blueshift and the symmetry of the features suggest they originate from a circular component situated beyond the farthest point of the eccentric disk (43.75 $\rm R_*$, see Figure \ref{fig:redshift_grav}). We thus experimented with adding the absorption produced by an additional gas ring  with a radial width of 0.5 $\rm R_*$ and with radii between 44 and 70 $\rm R_*$ (well inside the semi-major axis of the main transiting bodies). We assumed, for simplicity, the same vertical density structure as the one for the eccentric rings, and we computed the central densities from the radial scale in respect to the inner value of the disk, which gives values between $8 \times 10^{-7}$ and $3 \times 10^{-7}$ $\rm g/cm^3$ respectively. We also tested several temperatures (assumed constant throughout the ring) similar to those we found for the eccentric disk. Note that these ring properties are just for exploratory purposes and are certainly not unique (there is not much to be gained at this point by adding another layer of free parameters).

\begin{figure*}
    \centering{
    \includegraphics[width=0.9\textwidth]{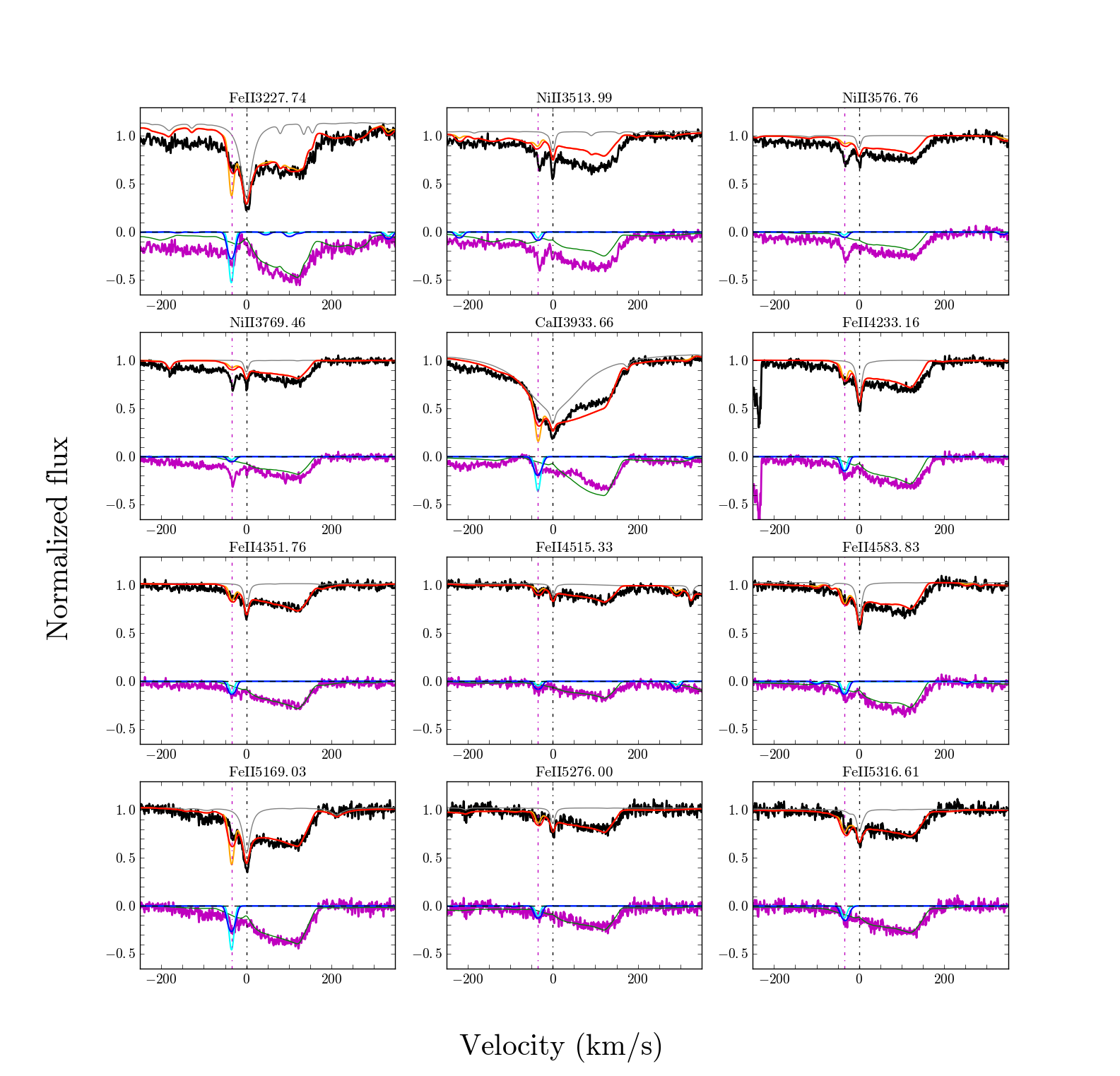}
    \caption{ Several regions that display the zero velocity/circular component for the 2016.04.01 data. Top of each panel shows the observed spectrum (black), the photospheric model (gray) and the disk model with a supplementary circular ring at 44 $\rm R_*$ (red) and 70 $\rm R_*$ (orange). The bottom of each panel shows the data from which we subtracted the photospheric contribution (magenta). The eccentric disk component is shown in green while the circular ring components for 44 $\rm R_*$ and 70 $\rm R_*$ are shown in blue and cyan, respectively.
	}}
    \label{fig:article_montrercoches}
\end{figure*}

Examples of the resulting absorption for two 8050 K rings situated respectively at 44 and 70 $\rm R_*$ are shown in Figure  \ref{fig:article_montrercoches}. Both rings produce features at the desired velocity shift from the photospheric line, but the ring farthest from the star produces, as expected, narrower lines, which seem to best reproduce the observations. The closer ring, however, seems to better represent the depth of the features (except for for Ni that is not well reproduce by either models) but given this is a simple constant temperature model, we refrain from drawing any definitive conclusion at this point.

\begin{figure}
    \centering{
    \includegraphics[width=\columnwidth]{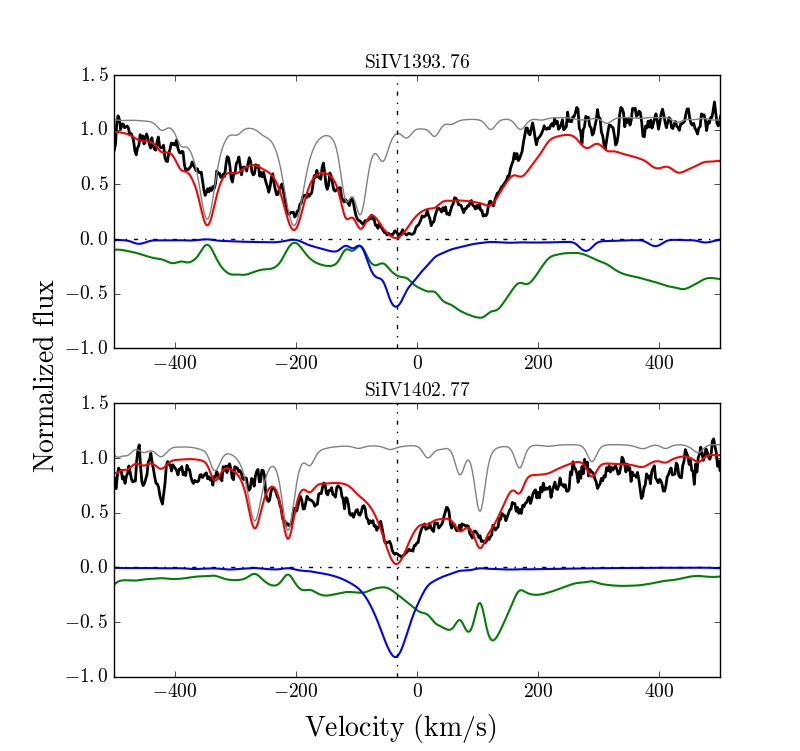}
    \caption{Circular + eccentric disk model in the Si IV region (2016.03.28 data). The circular ring has a constant temperature of 19,000 K, central density of $1 \times 10^{-6}$ $\rm g/cm^3$ and is situated at 44 $\rm R_*$. The blue and green lines represent the circular and eccentric disk components, respectively, the grey line the photosphere flux and the red line the combination of the three.
	}}
    \label{fig:article_SiIVring}
\end{figure}

A similar circular ring model can also be used to reproduce the zero velocity Si IV component observed in the UV (1393.76 and 1402.77 $\AA$ lines, see bottom right panel of Figure \ref{fig:article_raiesbien13000}). However, a much higher temperature is needed to produce transitions from this highly ionized species. Figure \ref{fig:article_SiIVring} shows the result from a constant 19,000 K ring model with a central density of $1 \times 10^{-6}$ $\rm g/cm^3$ situated 44 $\rm R_*$ from the star (presented here with the constant 13,000 K eccentric disk model which was doing a better job at reproducing the shifted Si IV features, see Figure \ref{fig:article_raiesbien13000}). While these parameters seem to reproduce the width and depth quite nicely, it is difficult to explain the presence of such a hot ring further away than the cooler eccentric disk. More elaborate ring models are clearly needed to fully understand the exact nature of those absorption features.

\subsection{Precession period}
\label{subsec:Precession period}

Now equipped with an eccentric ring model that can convincingly reproduce most of the circumstellar features present in the HIRES and HST data (2016.04.01 and 2016.03.28, respectively), we next try to apply it to reproduce circumstellar features present in spectroscopic data taken at other epochs. 

\begin{figure}
    \centering{
    \includegraphics[width=\columnwidth]{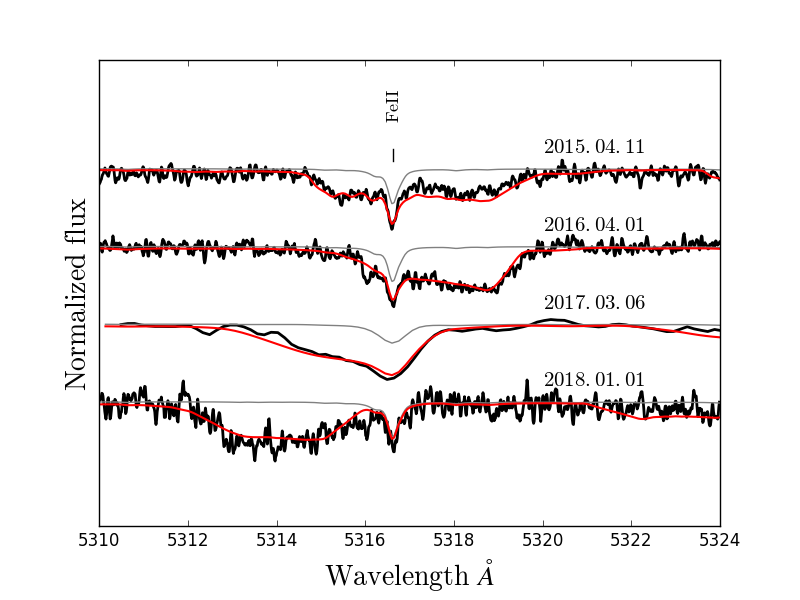}
    \caption{Best match in the precession cycle for the Fe II $5316.6$ at 4 different epochs. Photospheric synthetic spectrum is in gray and the model including gas disk absorption is in red.  
	}}
    \label{fig:article_bonnesepoques}
\end{figure}

\citet{Cauley2018} first estimated the precession period of the disk to be $\approx 5.3$ yrs, a value that we wish to revisit now that we have data covering a larger part of the cycle. We first try to find the position in the precession cycle that best reproduces the circumstellar features for each epoch of observation. This is done by rotating the geometric configuration presented in Figure \ref{fig:config_geo_t0}, and by calculating the corresponding absorption from the disk, until the shape of the circumstellar features are optimally reproduced. Of the remaining epochs of data available \citep[see Table 3 of][]{Xu2019}, we find there are 3 other times where the quality of the fit is particularly good and similar to that presented in Figures \ref{fig:article_raiesmauvaisesstructuretemp} and \ref{fig:figure8_struc_20160401}, namely 2015.04.11, 2017.03.06 and 2018.01.01 (note that epochs very near these dates are also acceptable). These four epochs are roughly 1 year apart and cover a good fraction of a whole cycle. Figure \ref{fig:article_bonnesepoques} shows a zoom on the $5316.6$ Fe II line for these dates (Figures similar to Fig. \ref{fig:article_raiesmauvaisesstructuretemp} and \ref{fig:figure8_struc_20160401} at these dates are presented in Appendix I).

\begin{figure}
    \centering{
    \includegraphics[width=\columnwidth]{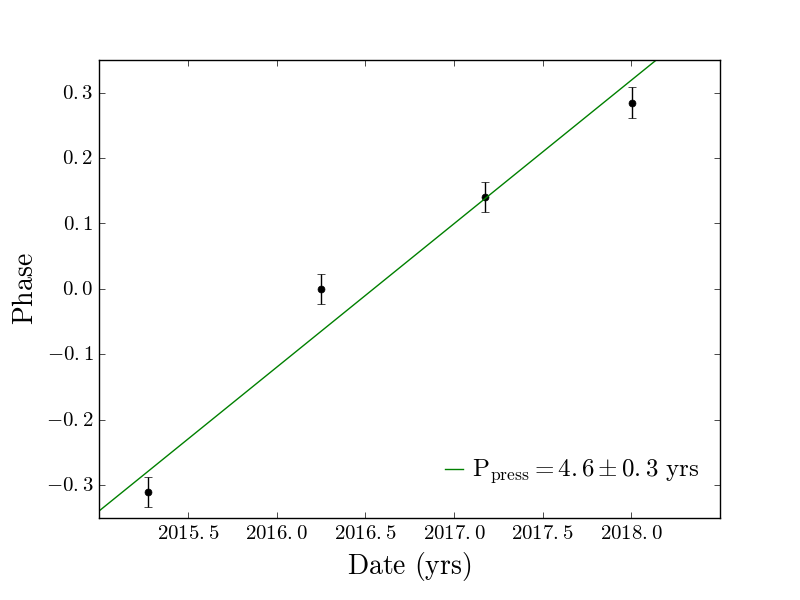}
    \caption{ Linear fit of the phases (green) for our best modeled epochs (black dots).
	}}
    \label{fig:article_fitperiode}
\end{figure}

From the angle of rotation (relative to the 2016.04.01, our zero point) needed to reach the configuration on the cycle that best matches the observed velocity profiles, we determine the corresponding phase at each epochs. Assuming that the precession rate is constant, we can perform a linear fit to the phases and obtain the precession period of the disk (see Figure \ref{fig:article_fitperiode}). We find a precession period of $4.6\pm0.3$ yrs, a slightly shorter value than the one estimated by \citet{Cauley2018}. It is interesting to note that the best fit with our assumed configuration is obtained, as in \citet{Cauley2018}, for a retrograde precession while GR-driven precession should be prograde. It is possible that there exists a different configuration that would better fit with prograde precession or that some disk pressure terms indeed contribute to produce a retrograde precession \citep[see][]{Miranda2018}, but discriminating between these posibilities is beyond the scope of this paper.

\begin{figure*}
    \centering{
    \includegraphics[width=0.7\textwidth]{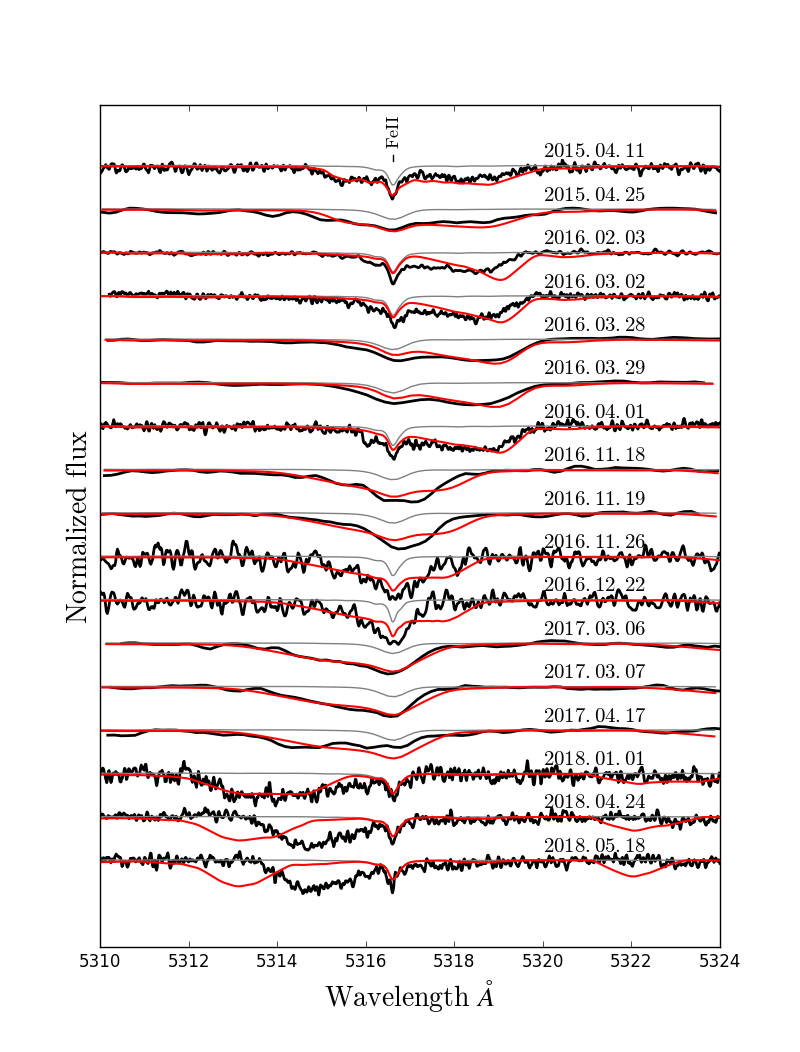}
    \caption{Circumstellar absorption model for the $5316.6$ $\AA$ Fe II line for each 17 epochs of data available assuming a precession cycle of 4.6 yrs. The cycle has been anchored to the fitted zero phase point in Figure  \ref{fig:article_fitperiode}. Photospheric model spectrum is in gray while the combined eccentric disk + photosphere is in red.
	}}
    \label{fig:article_period46an}
\end{figure*}

Figure \ref{fig:article_period46an} shows the prediction of our eccentric disk model in the region of the Fe II $5316.6$ line, a well isolated feature that is covered by all spectroscopic data, for all 17 epochs using this fitted period and zero point. In general, the model reproduces the overall shape and velocity spread for most of the epochs, although there are admittedly times where the the match is not perfect, or not good at all. In fact, there are some epochs for which there is no angle anywhere on the cycle that can reproduce the observed velocity profile.

\begin{figure}
    \centering{
    \includegraphics[width=\columnwidth]{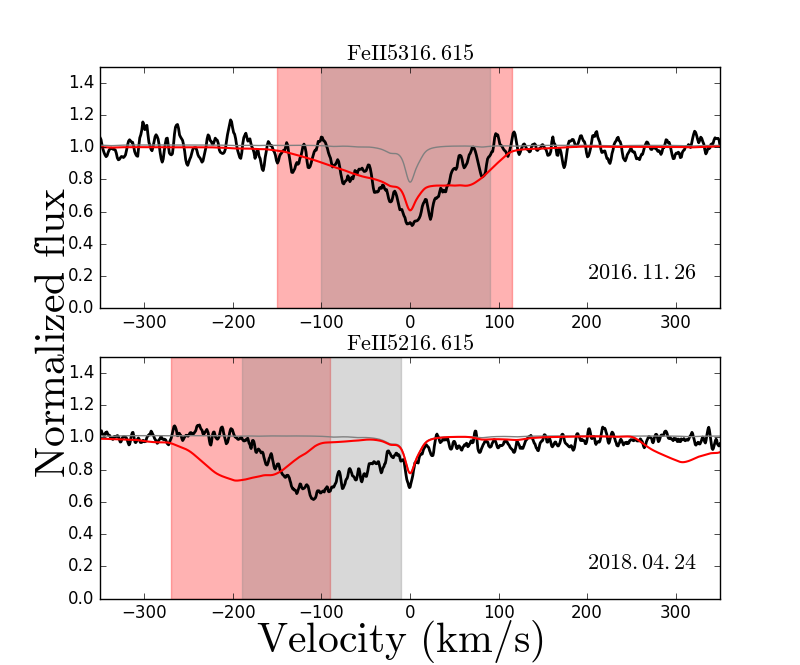}
    \caption{Two epochs of observation where the model is unable to reproduce the velocity profiles at any time in the cycle. Red line shows our best attempt to model the shape of the circumstellar features. The red and gray shaded areas correspond to the velocity spans of the model and observed features respectively.
	}}
    \label{fig:article_2016112620180424vitesse}
\end{figure}

In particular data taken in November and December 2016 seem to have narrower and deeper features and are almost centered on the photospheric line, which can never be reproduce by our model. Figure \ref{fig:article_2016112620180424vitesse} shows that the velocity width for the 2016.11.26 data is around 190 km/s and the model produces a wider feature of around 265 km/s. The April and May 2018 epochs are also problematic, as the features appear to have returned toward lower velocity shift much faster than predicted by the 4.6 yr period model. Note the model can produce narrow and deep features similar to those observed, but only at much greater velocities than observed (profiles produced at times where the velocity shift is lower are much wider, see Figure \ref{fig:article_precessioncycle}). 

While it is possible that the assumed constant precession rate is at fault here, these shortcomings are most probably an indication that the real geometric configuration of the eccentric disk is actually different than the one we are exploring in this paper. It is likely that some parameters that describe this configuration, namely the perihelion distances of the rings, eccentricities and angular shift of the apsidal line between the different rings, need to be tweaked a bit in order to achieve better results (for example, narrow features at lower velocities can probably be obtained with a smaller angular shift between rings that would be a bit farther from the star). Finding the exact configuration in this manner, however, is a very time consuming task that is outside the scope of this paper. A better approach left for future studies would probably be to invert the problem and infer the disk structure from the observed velocity profiles \citep[for example, see][]{Manser2016}.

\subsection{Transit spectra}
\label{subsec:Transit spectra}

\begin{figure}
    \centering{
    \includegraphics[width=\columnwidth]{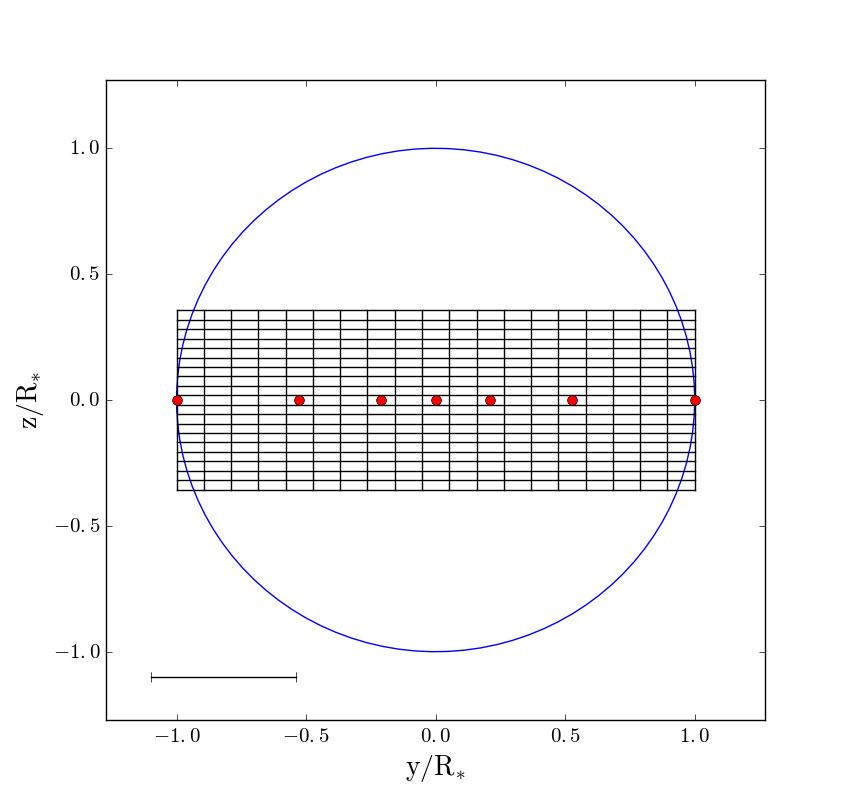}
    \caption{Position of the center of a blocking body (red dot) at 7 different positions during the transit (t1 to t7 from left to right). The grid corresponds to the surface covered by the disk. The error bar in the lower left corner corresponds to 5000 km.
	}}
    \label{fig:article_grilletransit}
\end{figure}

Simultaneous ground based photometric and spectroscopic observations of WD 1145+017 were secured on March 28, 2016 (two transits were covered with the 61'-inch telescope in Arizona while, at the same time, multiple 10 minutes sub-exposures were obtained with the Keck/ESI spectrograph). One interesting characteristic that was uncovered by these observations was that the circumstellar features were much shallower during the transit than out of transit \citep[see][for details]{Xu2019}. Using our simple eccentric model, it is possible to explain, in a semi-quantitative way, the behavior of the absorption features by simply blocking completely certain areas before integrating over the stellar surface, mimicking the passage of the asteroid (or disintegrating chunks detached from it) on our line of sight. For simplicity, we consider circular blocking objects (consisting of the solid disintegrating rocky body +  optically opaque gas cloud) situated at $\sim$ 101 $\rm R_*$ (Keplerian distance for a 4.5h orbit around a $0.656 \: M_{\odot}$ star) with diameters of 5000, 11,250 and 14,700 km. We also assume that the objects orbit in the same plane as the gas disk. For each object size, we compute the expected spectrum for 7 different positions during the transit (see Figure \ref{fig:article_grilletransit}).

\begin{figure}
    \centering{
    \includegraphics[width=0.8\columnwidth]{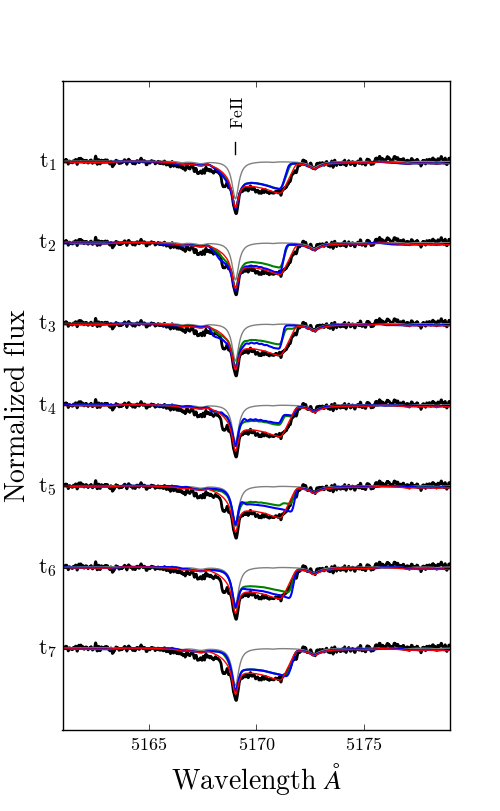}
    \caption{ Fe II 5169.03 line for the 7 positions of Figure  \ref{fig:article_grilletransit} for diameters of 11,250 km (green) and 14,700 km (blue). Photospheric model is in grey and the out of transit absorption is in red.
	}}
    \label{fig:article_transit100_150km}
\end{figure}

The first noteworthy result of this experiment is that the 5000 km transit, which blocks only 9\% of the surface of the star, barely has a detectable effect on the resulting spectrum at any point during the transit. We find that in order to have a noticeable effect, the blocking object has to be big enough to block not only a large part of the disk, but also regions of the star not covered by the disk. We estimate that the diameter thus has to be at least $\sim 6900$ km for a detectable depth variation during a transit.  Figure \ref{fig:article_transit100_150km} shows the resulting spectrum in the region of the 5169.03 Fe II line at the different moments during the transit for diameters of 11,250 and 14,700 km, blocking respectively 45\% and 80\% of the surface of the star.

\begin{figure*}
    \centering{
    \includegraphics[width=0.9\textwidth]{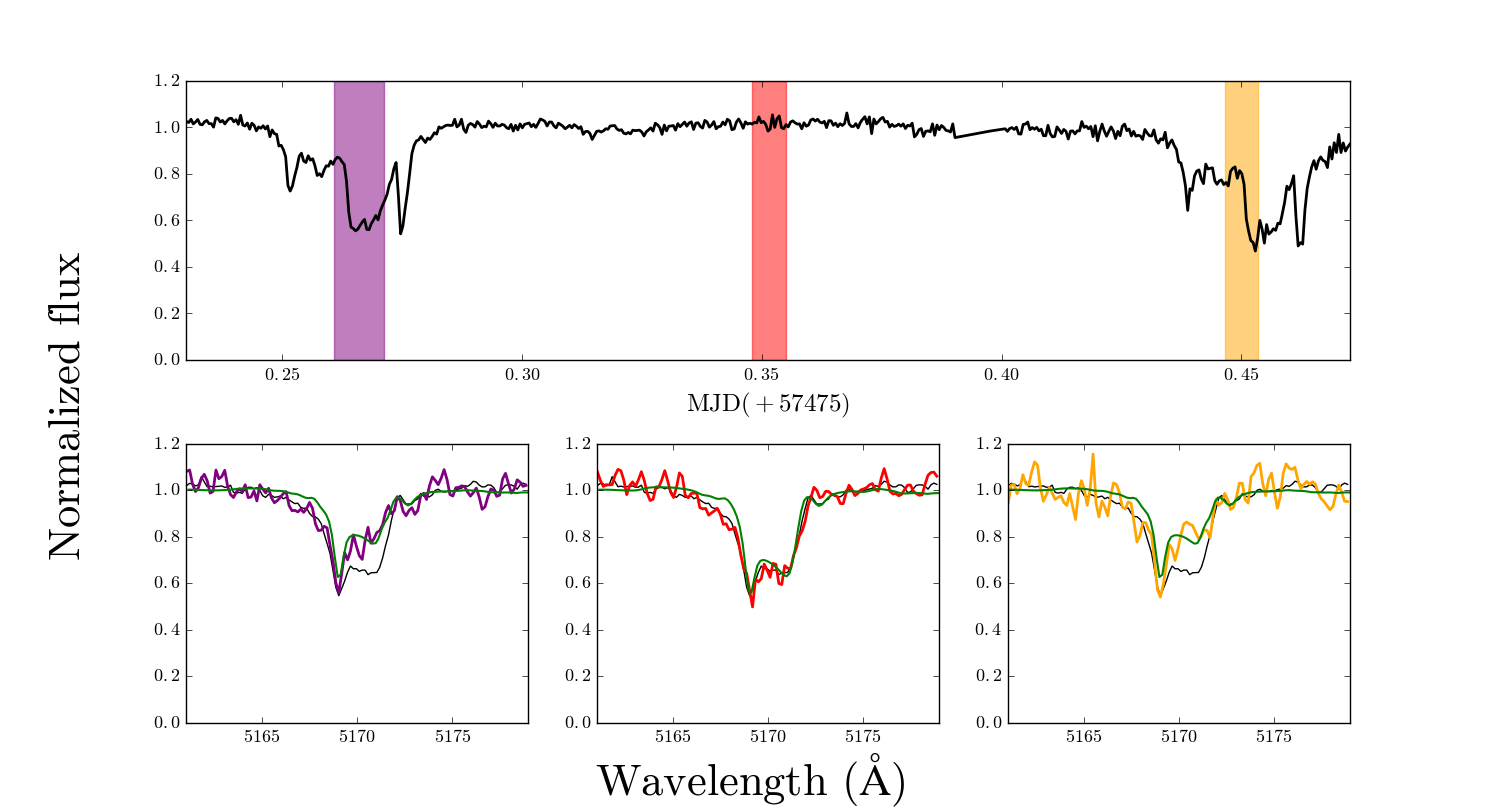}
    \caption{ Predicted and observed shape of the FeII 5169 absorption feature in and out of transit. Top: 61'-inch light curve obtained on 2016.03.28 (black). Bottom: subexposure of the ESI spectra for the three moments identified with the shaded areas in the top panel. Presented in black are the mean spectra outside the transit while the green line represents the predicted shape of the circumstellar feature according to our model at time $\rm t_4$ (first and third panel) and out of transit (middle panel).
	}}
    \label{fig:article_transitdata_both}
\end{figure*}

First of all, we observe that the features are systematically deeper for the 14,700 km body transit. This can be explained by the fact that this larger body blocks, relative to the smaller one, a larger fraction of the photosphere that is not covered by the eccentric disk. Note that had the transiting object blocked only part of the star that has no part of the eccentric disk in the line of sight, then the circumstellar features would have been deeper than when out of transit. Since the absorption features become shallower during transits \citep{Xu2019}, this confirms that the disintegrating asteroid and the disk are orbiting in the same plane, as suspected (if the orbit of the blocking body was inclined in relation to the eccentric disk, the circumstellar features would first become deeper, then shallower and then deeper again before returning to normal). It is also interesting to note that the shape of the features evolves during the transit, the most red-shifted side being more affected in the beginning of the transit while the lower velocity part is affected towards the end of the transit. This is a behavior that is expected given the line of sight velocities across the disk at this particular time (see Figure \ref{fig:velocity_profile_t0}). Unfortunately, the time resolution and signal-to-noise ratio from our March 28, 2016 observations are insufficient to detect such minute changes during the transit. Nevertheless, our simple model is able to successfully produce a reduction of the depth of the absorption features similar to those measured during the 2016.03.28 transits. Given that the light curve shows $\sim 40\%$ dips during transits, we can also rule out a body as large as 14,700 km since it would cover 80\% of the star surface (note there is the very likely possibility, however, that the transiting body is not totally optically thick, in which case it could indeed be much larger). Taking our simple 11,250 km diameter blocking area, we can compare the resulting shape of the circumstellar features at various points during the transit.

Figure \ref{fig:article_transitdata_both} shows the predicted shape of the Fe II 5169 region approximately at the maximum of the transit (corresponding to t4 in Figure \ref{fig:article_transit100_150km}) as well as when out of transit. While the temporal resolution and signal-to-noise of the sub-exposures are not sufficient to allow a better characterization of the transiting body, it is encouraging that this simple model is sensitive to details of the physical characteristics of the transiting object. The model also predicts, as observed, that the change in depths of the various absorption features are not the same for every transition, which is a reflection of the fact that they are not all formed at the same place in the disk. Moreover, it also naturally explains the shallower transits observed in the UV \citep[see][for a semi-analytical explanation of this phenomenon]{Xu2019}. In a forthcoming publication, we will attempt to extract a more detailed comprehension of the physical conditions in the disk based on the variation in depth of the various transitions during transiting events.

\section{Conclusion}
\label{sec:conclusion}

This paper first presented an updated analysis, based on new atmospheric parameters, of the photospheric composition of WD 1145+017. In particular, we now find that the somewhat high oxygen abundance problem reported in a previous study \citep{Xu2016} disappeared thanks in part to new high resolution spectroscopic data showing that some of the lines previously used for the abundance determination were contaminated by circumstellar absorption features. Our updated analysis now shows that the chemical abundance pattern observed at the photosphere is, to the first order, very close to what is expected for the accretion of a rocky body with bulk Earth composition.

We next developed an eccentric precessing disk model similar to that proposed by \citet{Cauley2018} in order to reproduce the numerous variable circumstellar features observed over time in the spectra of WD 1145+017. One of the main advantages of this model in comparison to that of \citet{Cauley2018} is that it predicts the shape of all circumstellar features from the UV to the optical using opacity calculations based on the assumed physical conditions of the disk. While the considered physical structure and dynamical configuration are certainly not a perfect representation of reality (in particular the very approximate temperature structure assumed), this simple model is able to reproduce the majority of the circumstellar features, from UV to optical, at numerous epochs during the precession cycle. Based on our calculations, the abundances for the circumstellar gas also appear to be about the same as those determined from the photosphere (no adjustment is needed to reproduce the relative depths from different elements). We also estimate that the total mass of the gas present in those rings is on the order of $\sim 10^{16}$ g, close to the mass of Uranus and Neptune's ring systems.

This eccentric disk model for circumstellar absorption also highlights the need for supplementary components, as some features cannot be accounted for by it. In particular, the numerous symmetrical zero velocity features (relative to the gravitationally red-shifted photospheric lines) observed in the optical probably indicate the presence of a supplementary low eccentricity ring outside the eccentric ring system. Furthermore, an additional low eccentricity component of hot gas is also needed to explain the presence of highly ionized species such as Si IV in the UV.

We find that a precession period of $4.6\pm0.3$ yrs is able to reproduce relatively well the observed shape of the velocity profile from April 2015 to January 2018, although some epochs, notably Nov/Dec 2016 and Apr/May 2018, are more challenging. Minor adjustments on the disk parameters (perihelion distances of the rings, eccentricities and angular shift of the apsidal line between the different rings) could probably provide an even better representation of the circumstellar features, a task left for future studies.

Finally, our simple model can also be used to confirm in a more quantitative way the semi-analytical description presented in \citet{Xu2019}, that is the anti-correlation between the circumstellar line strength and the transit depth, the shallower transit in the UV and the alignment of the orbital plane for the gas disk and the blocking fragment.

To conclude, considering the numerous approximations used in the construction of this eccentric disk model, its ability to reproduce a fair number of the observed characteristics of this system over time is quite satisfying. Future studies will aim at improving the physical structure of the rings (in particular implement a more realistic temperature structure) in order to better characterize the various components in orbit around WD 1145+017. A full analysis of all spectroscopic data available, including multiple epochs of HST data, will be presented elsewhere once the model has been improved and refined to a more satisfactory level.

\acknowledgments

This work was supported in part by NSERC (Canada) and the Fund FRQNT (Qu\'ebec).

\bibliographystyle{aasjournal}
\bibliography{references}

\section{Appendix I}

\begin{figure*}[h]
    \centering{
    \includegraphics[width=0.87\textwidth]{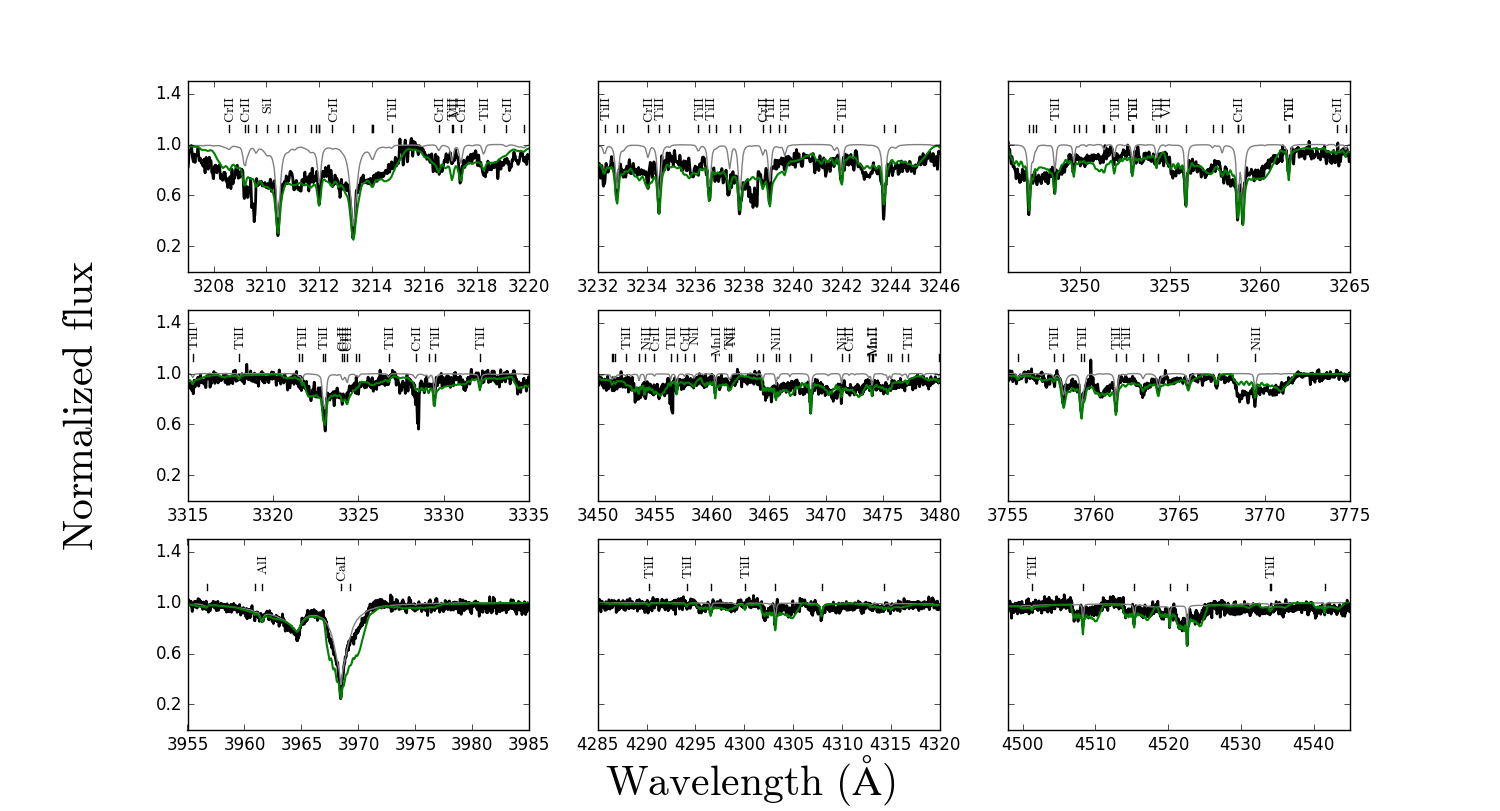}
    \caption{ Same regions presented in Figure \ref{fig:article_raiesmauvaisesstructuretemp} for the 2015.04.11 HIRESb epoch.
	}}
    \label{fig:figure11_20150411}
\end{figure*}

\begin{figure*}
    \centering{
    \includegraphics[width=\textwidth]{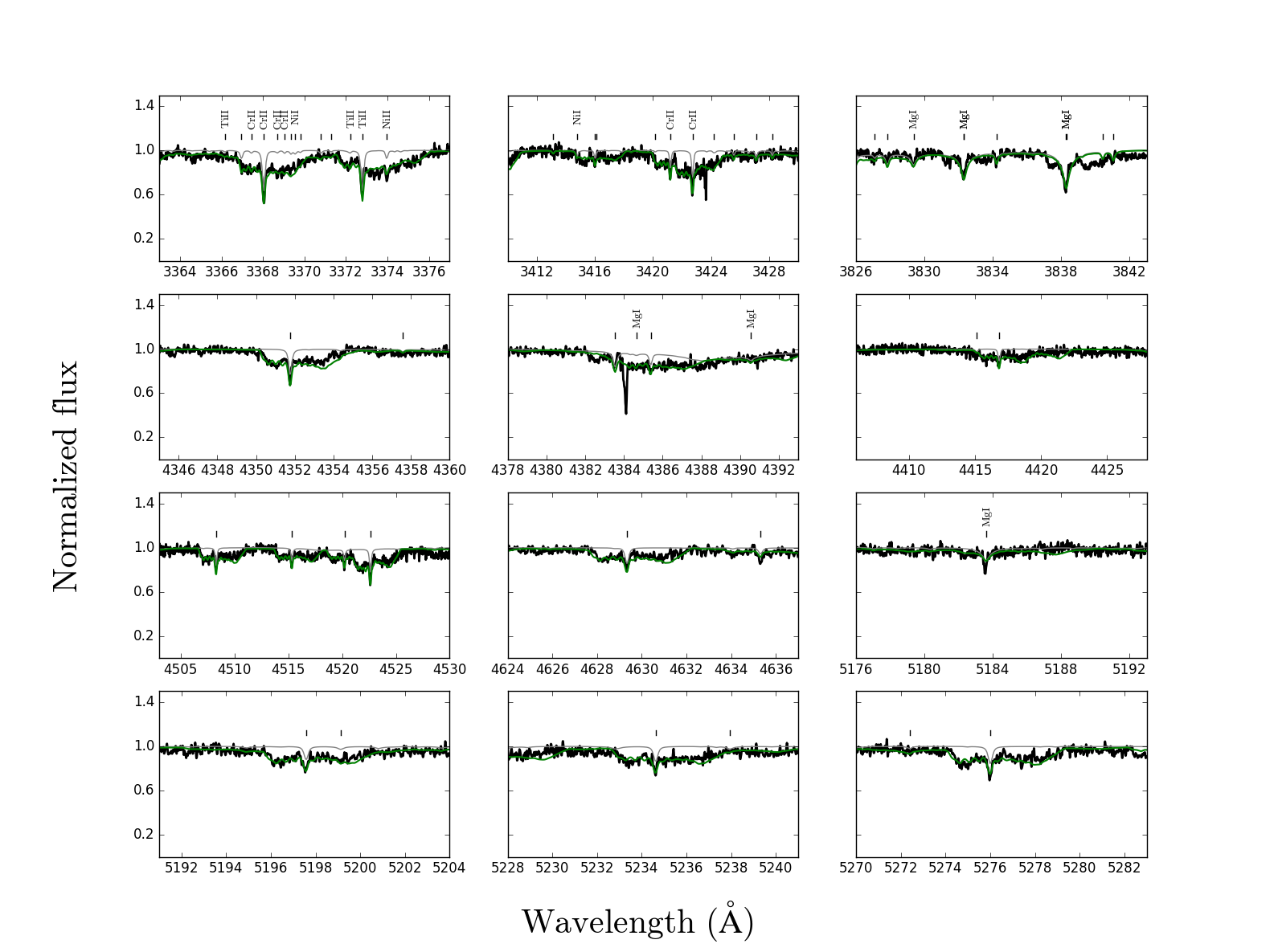}
    \caption{ Same regions presented in Figure \ref{fig:figure8_struc_20160401} for the 2015.04.11 HIRESb epoch.
	}}
    \label{fig:figure8_struc_20150411}
\end{figure*}

\begin{figure*}
    \centering{
    \includegraphics[width=0.87\textwidth]{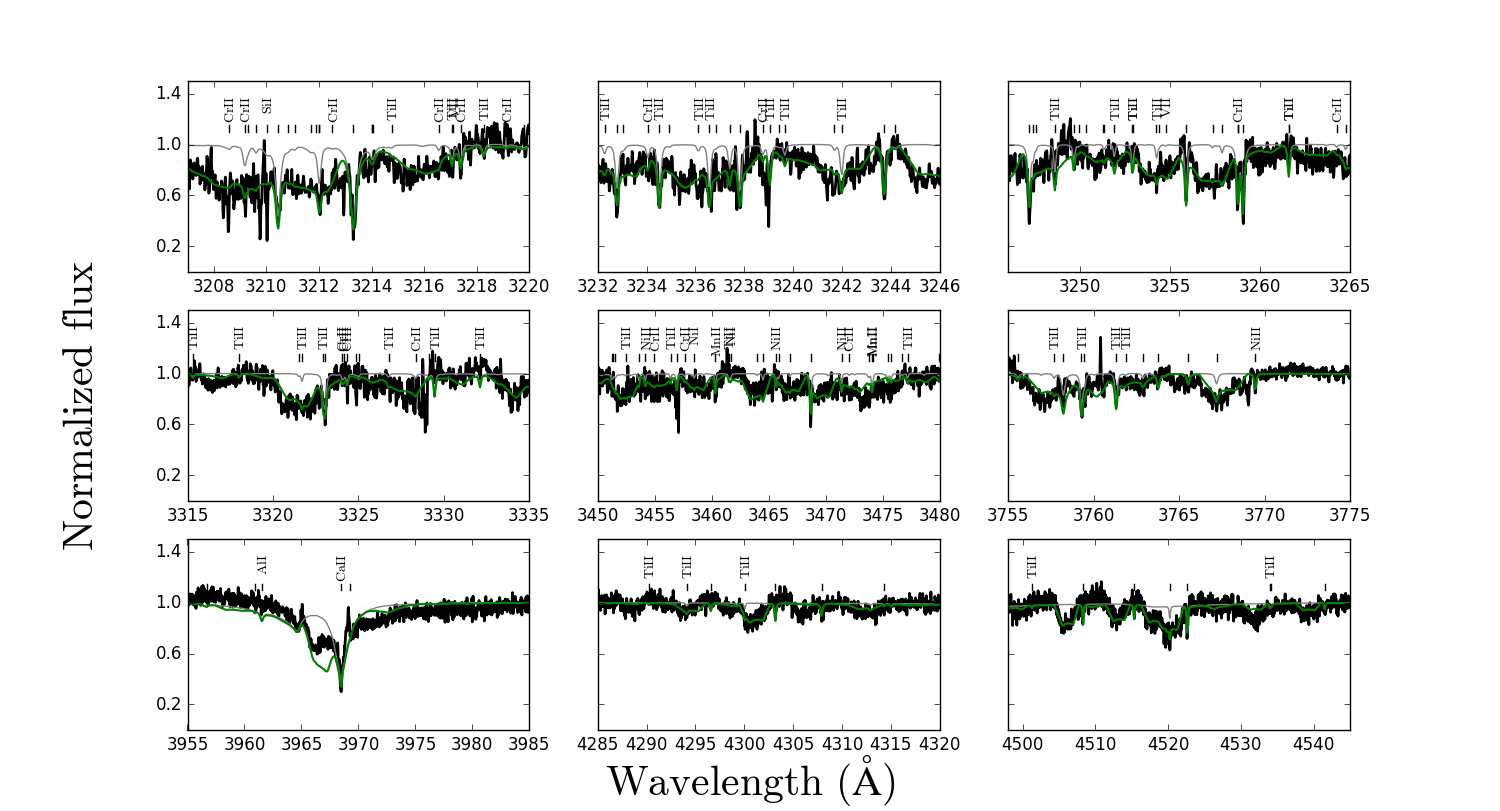}
    \caption{Same regions presented in Figure \ref{fig:article_raiesmauvaisesstructuretemp} for the 2018.01.01 HIRESb epoch.
	}}
    \label{fig:figure11_20180101}
\end{figure*}

\begin{figure*}
    \centering{
    \includegraphics[width=\textwidth]{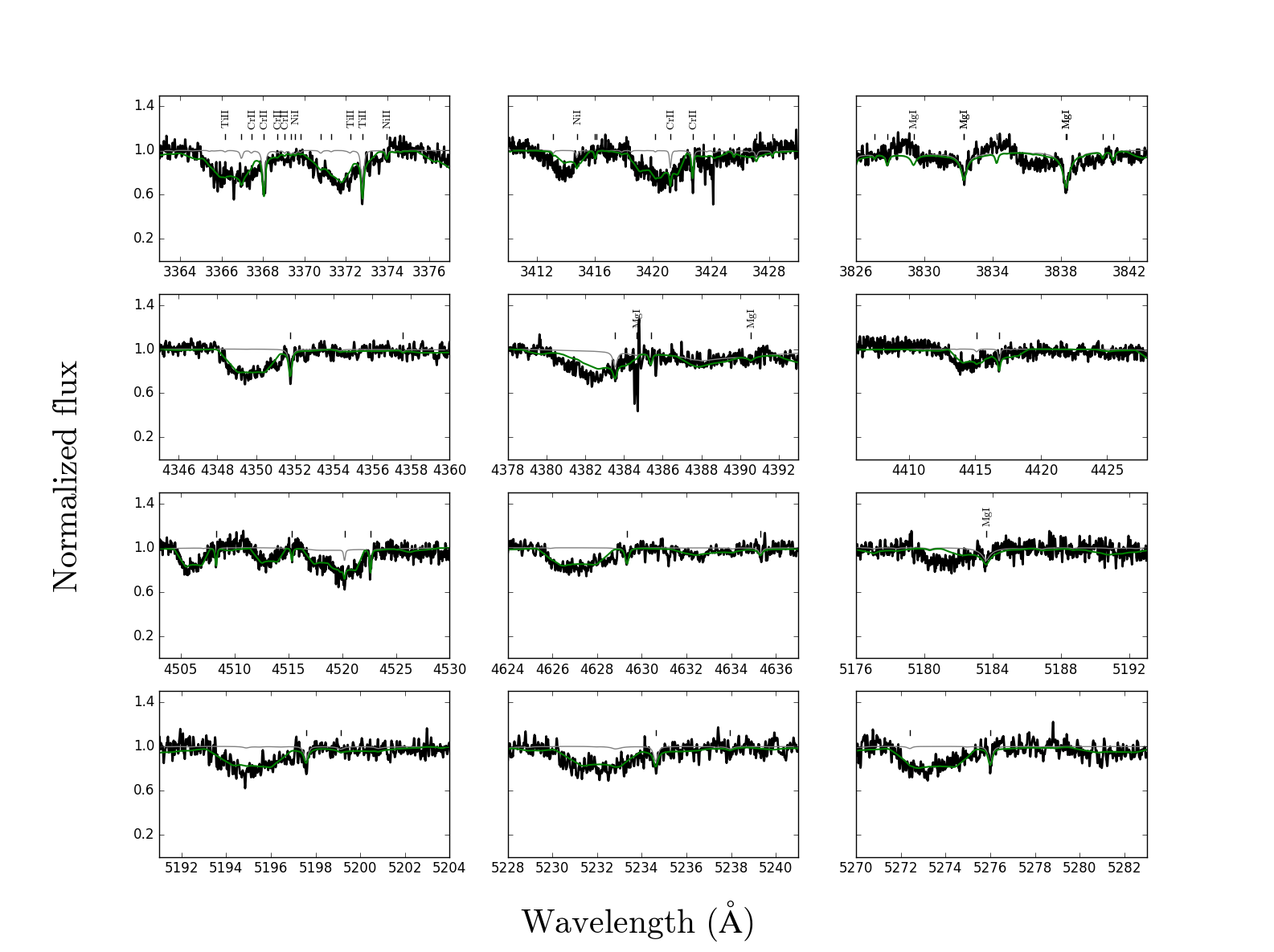}
    \caption{ Same regions presented in Figure \ref{fig:figure8_struc_20160401} for the 2018.01.01 HIRESb epoch.
	}}
    \label{fig:figure8_struc_20180101}
\end{figure*}
\end{CJK}

\end{document}